\documentclass[11pt]{article}

\usepackage[OT2,OT1]{fontenc}

\usepackage{a4wide}
\usepackage{amsmath}

\usepackage{mathrsfs}
\usepackage[T1]{fontenc}
\usepackage{mathpazo}
\usepackage{setspace}
\usepackage{amsfonts}
\usepackage{amssymb}
\usepackage{amsmath}
\usepackage{epsfig}
\usepackage{latexsym}
\usepackage{color}
\usepackage{nicefrac}
\usepackage[latin1]{inputenc}
\usepackage{cite}

\numberwithin{equation}{section}

\definecolor{papercolor}{rgb}{0.25, 0 ,1}
\definecolor{markcolor}{rgb}{1, 0.2, 0.2}

\definecolor{fux}{rgb}{0.75, 0 ,1}

\author{
  \begin{minipage}{.97\linewidth}
    \vspace{1cm}
       \begin{center}
      \begin{small}
        \textbf{Marco M. Caldarelli}$^{1,2}$, 
                \textbf{Robert G. Leigh}$^3$,  
              \textbf{Anastasios C. Petkou}$^4$, 
                      \textbf{P. Marios Petropoulos}$^1$, 
        \textbf{Valentina Pozzoli}$^1$ and
                       \textbf{Konstadinos Siampos}$^1$        
              \end{small}
    \end{center}
    \vspace{0.5cm}
    \hspace{2cm}\begin{minipage}{.7\linewidth}
\begin{center}     {\it \begin{footnotesize}
\hbox{\kern-1cm\vbox{\vskip0cm
 \begin{itemize}
              \item[$^1$] Centre de Physique Th\'eorique\\ 
        Ecole Polytechnique, CNRS UMR 7644\\
        91128 Palaiseau Cedex, France\\
{\tt caldarelli, marios, pozzoli, \\ ksiampos@cpht.polytechnique.fr}
 \item[$^2$] Laboratoire de Physique Th\'eorique\\ 
  Universit\'e de Paris-Sud 11,   B\^at. 210\\ 
  CNRS UMR 8627\\
  91405 Orsay Cedex, France
\end{itemize}}
\kern-5cm\vbox{
\begin{itemize}
         \item[$^3$] Department of Physics, University of Illinois\\1110 W. Green Street\\ Urbana IL 61801, U.S.A.\\
{\tt rgleigh@illinois.edu}  
\vskip0.47cm
  \item[$^4$] Department of Physics, University of Crete\\ 71003 Heraklion, Greece\\
      {\tt petkou@physics.uoc.gr}
 \end{itemize}\vskip0.05cm
}}
     \end{footnotesize}}
\end{center}
    \end{minipage}
    \vspace{0.5cm}
  \end{minipage}
}

\title{\vspace{1.5cm}
 \boldmath \begin{huge}
    \textbf{Vorticity in holographic fluids}
  \end{huge} \unboldmath
}

\begin{document}

\begin{titlepage}
  \maketitle
  \thispagestyle{empty}

  \vspace{-12.5cm}
  \begin{flushright}
    CPHT-RR022.0512\\
LPT-ORSAY 12-60\\
  \end{flushright}

  \vspace{10cm}

  \begin{center}
    \textsc{Abstract}\\
  \end{center}
In view of the recent interest in reproducing holographically          
various properties of conformal fluids, we  review the issue of         
vorticity in the context of AdS/CFT. Three-dimensional fluids with vorticity require four-dimensional bulk geometries with either angular momentum or nut charge,
 whose boundary geometries fall into the Papapetrou--Randers class. The boundary fluids emerge in stationary non-dissipative kinematic configurations,
 which can be cyclonic or vortex flows, evolving in compact or non-compact supports.  A rich network of Einstein's solutions arises, naturally connected 
with three-dimensional Bianchi spaces. We use Fefferman--Graham expansion to handle holographic data from the bulk and discuss the alternative for 
reversing the process and reconstruct the exact bulk geometries.

\vspace{2cm} \noindent Review to appear in the proceedings of the
\textsl{Corfu Summer Institute 2011: School and workshops on elementary particle physics and gravity,}	September 4--18, 2011,	Corfu, Greece.

\end{titlepage}

\onehalfspace

\tableofcontents

\section{Introduction}
The relationship of fluid dynamics with general relativity goes back to the work of Damour \cite{Damour:1979}.  Lately, a different and perhaps more concrete shape of this relationship has been given by the so-called fluid/gravity correspondence (see for example the recent review \cite{Hubeny}). According to the latter, the gravitational degrees of freedom that reside in the boundary of an asymptotically AdS$_D$ spacetime describe the hydrodynamics of a relativistic fluid in $D-1$ dimensions. Consequently, the dynamical equations of the latter systems (e.g.~Euler or Navier--Stokes)  are encoded in the asymptotic behaviour of the bulk Einstein equations. 

The fluid/gravity correspondence framework appears to be capable of describing different facets of relativistic fluids, such as superfluidity and dissipation. This appears to be a novel mechanism of emergence in physics whereby the low-energy effective degrees of freedom arise holographically in the boundary of a gravitational system.   At a practical level, fluid/gravity correspondence currently occupies a large part of the AdS/CMT correspondence as are collectively called the efforts to find new computational tools for strongly coupled condensed matter systems using holography (see e.g.~the reviews \cite{Hartnoll:2009sz,Herzog:2009xv,Horowitz:2010gk}). 

Despite some important results in the study of holographic fluids, the issue of vorticity has been less well understood. This is important if one wants to extend the realm of AdS/CMT to interesting condensed matter systems such as rotating Bose or Fermi gases \cite{Cooper, Fetter}, turbulence or wave propagation in moving metamaterials (e.g. \cite{Leonhardt}). With such extensions in mind, we review here our recent  attempt to setup a holographic framework  for the description of fluids with vorticity. Even without touching the thorny question of dissipation, \emph{i.e.}
assuming  local equilibrium and non-dissipating kinematics, our studies reveal a remarkably rich structure as soon as vorticity is switched on. In particular, we note the intimate relationship of our neutral rotating holographic fluids with charged fluids in magnetic fields, as well as  with the  problem of wave propagation in moving media. We believe that the latter observation can lead to the holographic description of analogue gravity systems \cite{Unruh1, Unruh2,Barcelo,Liberati}.  

In the present review we choose to devote most of our discussion to the gravitational side of the duality and we extensively discuss in Sec. \ref{RF} relativistic fluids, in Sec. \ref{RP} the general stationary Papapetrou--Randers and Zermelo geometries and in Sec. \ref{ex21} various concrete examples of stationary geometries in $2+1$ dimensions. In Sec. \ref{holog} we review the Fefferman--Graham construction of holographic fluids in $2+1$ dimensions focusing on the Kerr--AdS$_4$, the Taub--NUT--AdS$_4$ and the hyperbolic NUT--AdS$_4$ solutions. In Sec. \ref{Min} we give an overview of some new results, to be presented elsewhere \cite{NewPaper}, that aim to connect our approach with alternative  descriptions of holographic fluids. Section  \ref{con} contains our conclusions.

\section{Relativistic fluids}\label{RF}

In this section, we recall the salient features of relativistic fluid dynamics (see e.g.~\cite{Ehlers:1993gf, vanElst:1996dr}). This includes aspects of vector-field congruences and properties of the energy--momentum tensor. We work in arbitrary spacetime dimension $D$.

\subsection{Vector-field congruences}\label{vfc}

We consider a manifold endowed with a spacetime metric of the generic form
\begin{equation}\label{Dmet}
\mathrm{d}s^2 =g_{\mu\nu}\mathrm{d}x^\mu \mathrm{d}x^\nu= \eta_{ab}\hat{e}^a \hat{e}^b.
\end{equation}
We will use $a,b,c,\ldots =0,1,\ldots, D-1$ for transverse Lorentz indices along with $\alpha,\beta,\gamma=1,\ldots, D-1$. 
Coordinate indices will be denoted $\mu,\nu,\rho, \ldots$ for spacetime $\mathrm{x}\equiv (t,x)$ and  $i,j,k, \ldots$ for 
spatial $x$ directions. The dual of the orthonormal coframe $\hat{e}^a$ is  the frame $\check{e}_a$, which 
satisfies $\hat{e}^a(\check{e}_b)=\delta^a_b$. To define parallel transport we take  the Levi--Civita connection coefficients $\Gamma^a_{bc}$ defined via the spin-connection one-form $\hat{\omega}^a_{\phantom{a}b}$ as
\begin{equation}
\label{LCconn}
\mathrm{d}\hat{e}^a+\hat{\omega}^{a}_{\hphantom{a}b}\wedge \hat{e}^b=0,\quad\hat{\omega}^a_{\hphantom{a}b}=\Gamma^a_{bc}\hat{e}^c, \quad\nabla_{\check{e}_a}\check{e}_b
=\Gamma^c_{ba}\check{e}_c.
\end{equation}

Consider now an arbitrary timelike vector field $\hat{u}=u_a\hat{e}^a$, normalized as  $\eta_{ab}u^au^b=-1$,  later identified with the fluid velocity. Its integral curves define a congruence which is characterized by its acceleration, shear, expansion and vorticity:
\begin{equation}
\label{def1}
\nabla_{a} u_b=-u_a a_b +\frac{1}{D-1}\Theta h_{ab}+\sigma_{ab} +\omega_{ab}
\end{equation}
with\footnote{Our conventions are: $A_{(ab)}=\nicefrac{1}{2}\left(A_{ab}+A_{ba}\right)$ and $A_{[ab]}=\nicefrac{1}{2}\left(A_{ab}-A_{ba}\right)$.}
\begin{eqnarray}
a_a&=&u^b\nabla_bu_a, \quad
\Theta=\nabla_a u^a, \label{def21}\\
\sigma_{ab}&=&\frac{1}{2} h_a^{\hphantom{a}c} h_b^{\hphantom{b}d}\left(
\nabla_c u_d+\nabla_d u_c
\right)-\frac{1}{D-1} h_{ab}h^{cd} \nabla_c u_d \label{def22}\\
&=& \nabla_{(a} u_{b)} + a_{(a} u_{b)} -\frac{1}{D-1} h_{ab} \nabla_c u^c ,
\label{def23}\\
\omega_{ab}&=&\frac{1}{2} h_a^{\hphantom{a}c} h_b^{\hphantom{b}d}\left(
\nabla_c u_d-\nabla_d u_c
\right)= \nabla_{[a} u_{b]} + u_{[a} a_{b]}.\label{def24}
\end{eqnarray}
The latter allows to define the vorticity form as
\begin{equation}\label{def3}
2\omega=\omega_{ab}\, \hat{e}^a\wedge\hat{e}^b =\mathrm{d}\hat{u} +
\hat{u} \wedge\hat{a}\, .
\end{equation}
These tensors satisfy several simple identities:
\begin{equation}
u^a a_a=0, \quad u^a \sigma_{ab}=0,\quad u^a \omega_{ab}=0, \quad u^a \nabla_b u_a=0, \quad h^c_{\hphantom{c}a} \nabla_b u_c =\nabla_b u_a.
\end{equation}

Killing vector fields, satisfying $\nabla_{(a}\xi_{b)}=0$, are congruences with remarkable properties. We quote two of them, the proof of which is straightforward:
\begin{itemize}
\item A Killing vector field has vanishing expansion. 
\item A constant-norm\footnote{This is not an empty statement since Killing vectors cannot be normalized at will. When their norm is constant, it can be consistently set to $-1, 0$ or $+1$.} Killing vector field is furthermore geodesic and shearless. It can only carry vorticity.
\end{itemize}

The timelike vector field $\check{u}$  can be used to decompose any tensor field on the manifold in transverse and longitudinal components with respect to itself.  The decomposition is performed by introducing the longitudinal and transverse projectors:
\begin{equation}
\label{proj}
U^a_{\hphantom{a}b} = - u^a u_b, \quad h^a_{\hphantom{a}b} =  u^a u_b + \delta^a_{b},
\end{equation}
where $h_{ab}$ is also the induced metric on the surface orthogonal  to $\check{u}$. The projectors satisfy the usual identities:
\begin{equation}
U^a_{\hphantom{a}c} U^c_{\hphantom{c}b} = U^a_{\hphantom{a}b},\quad U^a_{\hphantom{a}c} h^c_{\hphantom{c}b}  =   0 , \quad h^a_{\hphantom{a}c} h^c_{\hphantom{c}b}  =   h^a_{\hphantom{a}b} ,\quad U^a_{\hphantom{a}a}=1, \quad h^a_{\hphantom{a}a}=D-1.
\end{equation}
For example, any rank-two symmetric tensor $T_{ab}$ can be decomposed in longitudinal, transverse and mixed components: 
\begin{equation}\label{Tdec}
T_{ab}=e u_a u_b+ S_{ab}-u_a q_b - u_b q_a,
\end{equation}
the non-longitudinal part being 
\begin{equation}\label{Sdec}
\Sigma_{ab}=S_{ab}-u_a q_b - u_b q_a.
\end{equation}
We have defined
\begin{equation}
e= u^a u^b T_{ab},\quad
S_{ab}=  h_a^{\hphantom{a}c} h_b^{\hphantom{b}d} T_{cd},\quad
q_a= h_a^{\hphantom{a}b}T_{bc}u^c,
\end{equation}
such that 
\begin{equation}\label{trans}
u^a q_a=0, \quad u^a S_{ab}=0.
\end{equation}
Finally
\begin{equation}
u^a T_{ab}=q_b-e u_b.
\end{equation}

\subsection{The energy--momentum tensor}

The mere existence of a metric and a timelike vector-field congruence does not necessarily imply the presence of a relativistic fluid. If we wish to identify the timelike vector $\check{u}$ with the velocity of a relativistic fluid, then we should require the presence of an additional symmetric rank-two tensor field -- the energy--momentum tensor $T_{ab}$ whose projection along $\check{u}$ is the (positive) energy density $\varepsilon$ of the fluid
\begin{equation}
T_{ab}u^au^b=\varepsilon\,,
\end{equation}
as measured in the local proper frame. The latter concept deserves a comment. In non-relativistic fluids, the velocity field is unambiguously defined as the velocity of the mass flow of the fluid. In the relativistic case, it requires a more formal definition as energy and mass cannot be distinguished, and energy flows can be the result of dissipative phenomena or thermal conduction. One way to define the velocity, which amounts to defining a specific local proper frame known as \emph{Landau frame}, is to demand the absence of mixed terms in \eqref{Sdec}:
\begin{equation}\label{Stra}
 u^a \Sigma_{ab}=0.
\end{equation}

Let us continue applying the decomposition \eqref{Tdec} to the energy--momentum tensor. Inserting \eqref{Sdec} in \eqref{Stra}, Eqs.  \eqref{trans} imply that $q^a$ vanishes in the Landau frame, where the energy--momentum is thus
\begin{equation}\label{e-m}
T_{ab}=\varepsilon u_a u_b+ S_{ab}.
\end{equation}
The last piece $S_{ab}$ is the stress tensor, purely transverse. 

For a perfect fluid, all information is encapsulated in a further unique piece of data: the pressure $p$
measured in the local proper frame. Hence, the stress tensor reads:
\begin{equation}
\label{perstr}
 S_{ab}^{\mathrm{perf.}}=p  h_{ab}.
\end{equation}
For a viscous fluid, the stress tensor contains friction terms:
\begin{equation}\label{str}
 S_{ab}=p  h_{ab}+t_{ab},
\end{equation}
where $t_{ab}$ is usually expressed as an expansion in the derivatives of the velocity field.
At lowest order one finds
\begin{equation}\label{Sigvisc}
t_{ab}=-2\eta \sigma_{ab}-\zeta h_{ab}\Theta
\end{equation}
with $\eta, \zeta$ the shear and bulk viscosities. In $2+1$ dimensions there exists another term at this order, breaking the parity symmetry: $\zeta_{\mathrm{H}}\,  \epsilon^{\vphantom{c}}_{cd(a} u^c\sigma^d_{\hphantom{d}b)}$. The coefficient $\zeta_{\mathrm{H}} $ is the \emph{rotational Hall viscosity}. It characterizes a transport phenomenon similar to the Hall conductivity of charged fluids in magnetic fields.  

The dynamical equations for the fluid (Euler, Navier--Stokes, \dots)  are all encoded in the covariant conservation of the energy--momentum tensor 
\begin{equation}\label{cons}
\nabla^a T_{ab}=0\,.
\end{equation}
In the non-relativistic limit, Eq.~\eqref{cons} also delivers a matter-current conservation, which, for relativistic fluids, must be introduced separately as a consequence of charge conservation, if any. 

\subsection{Effectively perfect fluids}\label{vpf}
Relativistic fluids in the hydrodynamic regime are long wavelength approximations of finite-chemical-potential and finite-temperature states of certain (unknown) quantum field theories\footnote{A special class of such fluids, actually the one that naturally arises in holography,  are conformal fluids, \emph{i.e.}~those having vanishing energy--momentum trace: $(D-1)p-\varepsilon =(D-1)\zeta \Theta$. This equation is supposed to hold for any kinematic configuration, in particular when the fluid is at rest, where $\varepsilon=(D-1)p$. The latter is therefore adopted as a thermodynamic equation of state valid \emph{always} locally. When the fluid is not at rest, we conclude then that $\zeta \Theta=0$, which must hold for any $\Theta$. In this scheme, the bulk viscosity for a conformal fluid is thus vanishing identically. Similar conclusions are reached for higher-order viscosity coefficients entering the traceful part of the energy--momentum tensor.}.
Quite generically all such fluids exhibit dissipative phenomena as they describe media with non-zero shear viscosity. However, {\it all} such fluids can be in special kinematic configurations where the effects of dissipation are ignorable\footnote{A fluid can be stationary and altogether dissipate energy provided it is not isolated. These situations are better designated as \emph{forced steady states}. On curved boundary backgrounds, the forcing task can be met by gravity through the boundary conditions. This was discussed in \cite{Bhattacharyya:2008ji}. As this feature does not appear in the backgrounds that will be analyzed in the forthcoming sections, we will not pursue this further.\label{forced}}. In this case, their dynamics is captured by the perfect part of the stress tensor and the equations of motion read\footnote{The interested reader can find more information on these specific issues in e.g.~\cite{Caldarelli:2008mv, Caldarelli:2008ze}.}:
\begin{equation}
  \begin{cases}\label{perfeom}
  (\varepsilon+p)\Theta+\nabla_{\check{u}}\varepsilon=0
 \\ (\varepsilon+p)\hat{a}+\nabla_{\perp}p=0
\end{cases}
\end{equation}
($\nabla_{\perp}=\nabla+\hat{u}\nabla_{\check{u}}$ stands for the covariant derivative along the directions normal to the velocity field). 
Under this assumption, taking also into account the conformality ($\varepsilon= (D-1)p\propto T^D$), Eqs. \eqref{perfeom} lead to 
\begin{equation}
  \begin{cases}\label{perfeom-dissless}
 \nabla_{\check{u}}\varepsilon=0
 \\ \hat{a}=-\frac{\nabla_{\perp}p}{Dp}.
\end{cases}
\end{equation}
The energy density is conserved along the fluid lines and in the absence of spatial pressure gradients (\emph{i.e.}~for energy and pressure constants in spacetime), the flow is geodesic. 

In several instances, the velocity of the fluid turns out to be a Killing vector field. Then, from the discussion of Sec.~\ref{vfc} several straightforward conclusions can be drawn: 
\begin{itemize}
\item The flow is geodesic, shearless and expansionless.
\item The internal energy density is conserved and the pressure is spatially homogeneous.
\item If the fluid is conformal then $\varepsilon= (D-1)p\propto T^D$ is  constant in spacetime.
\end{itemize}
Therefore, despite its viscosity, the kinematic state of the fluid can be  steady and non-dissipa- tive. For this to happen, however, the existence of a constant-norm timelike Killing vector is required. In other words, the background geometry must itself be stationary\footnote{More generally, it can be shown that the velocity field $u^\mu$ of a stationary fluid flow has to be proportional to a Killing vector field of the background geometry \cite{Caldarelli:2008mv}.}. In this case, the constant-norm timelike Killing vector congruence allows for the definition of a global time coordinate, with associated inertial frames. The latter are comoving with the fluid. All the examples that we will discuss in the following fall into this class.

\section{Papapetrou--Randers stationary geometries}\label{RP}

Starting with appropriate time-independent bulk backgrounds\footnote{Notice that in some instances time independence is not met in the bulk, but stationarity remains valid on the boundary as a consequence of appropriate boundary conditions \cite{Bhattacharyya:2008ji}. These cases belong to the class of forced dissipative steady-states mentioned in footnote \ref{forced}.},  conformal fluids appear holographically, evolving generally on stationary but not necessarily static boundary geometries. Those fluids possess therefore non-dissipative dynamics inherited from the gravitational environment and this dynamics contains in general vorticity. We will present here some basic properties of the boundary backgrounds arising in this  context and explain how they affect the fluid dynamics. We postpone to Sec.~\ref{holog} the actual holographic analysis relating some of these backgrounds to exact bulk Einstein spacetimes.

\subsection{General properties, geodesic congruences and Papapetrou--Randers frame}
\label{Papasec}

Stationary metrics appearing in the holographic analysis we will be presenting later on are of the generic form
\begin{equation}
\label{Papa}
\mathrm{d}s^2=B^2\left(-(\mathrm{d}t-b_i \mathrm{d}x^i)^2+a_{ij}(x)\mathrm{d}x^i \mathrm{d}x^j\right),
\end{equation}
where $B, b_i, a_{ij}$ are $x$-dependent functions. These metrics were introduced by Papapetrou in \cite{Papapetrou}. 
They will be called hereafter \emph{Papapetrou--Randers}  because they are part of an interesting network of relationships 
involving the Randers form \cite{Randers}, discussed in detail in \cite{Gibbons} and more recently used in \cite{Leigh:2011au,LPP2}. 

For later convenience, we introduce  $a^{ij}, b^i$ and $ \gamma$
such that
\begin{equation}
a^{ij}a_{jk}= \delta^i_k,\quad  b^i=a^{ij}b_j,\quad     \gamma^2 = \frac{1}{1-a^{ij}b_i b_j}.
\end{equation}
The metric components read:
\begin{equation}
g_{00}=-B^2,\quad g_{0i}=B^2 b_i, \quad
g_{ij}=B^2(a_{ij}-b_ib_j),
\end{equation}
and those of the inverse metric:
\begin{equation}
g^{00}=-\frac{1}{\gamma^2B^2},\quad g^{0i}=\frac{b^i}{B^2}, \quad
g^{ij}=\frac{a^{ij}}{B^2}.
\end{equation}
Finally, 
\begin{equation}
\sqrt{-g}=B^D \sqrt{a},
\end{equation}
where $a$ is the determinant of the symmetric matrix with entries $a_{ij}$.

In the natural frame of the above coordinate system $\{\partial_t, \partial_i\}$, any observer at rest has normalized velocity $\check{u}=\frac{1}{B}\partial_t$ and dual form $\hat{u}=-B(\mathrm{d}t-b)$ (we set $b=b_i \mathrm{d}x^i$). The normalized vector field $\check{u}$ is not in general Killing -- as opposed to $\partial_t$. For this observer, the acceleration is thus non vanishing:
\begin{equation}\label{accRan}
\check{a}=\nabla_{\check{u}} \check{u}= g^{ij}\partial_i \ln B\left(\partial_j + b_j\partial_t\right).
\end{equation}
As already mentioned, the motion is inertial if and only if $B$ is constant. It will be enough for our purposes to consider the case $B=1$, and all subsequent formulas will assume this choice.  
We will furthermore introduce a frame
\begin{equation}
\label{frameRanders}
\check{e}_0 =  \partial_t ,\quad  \check{e}_\alpha=E_\alpha^{\hphantom{\alpha}i}\left(b_i\partial_t+\partial_i\right), \quad E_\alpha^{\hphantom{\alpha}i}E^\beta_{\hphantom{\beta}i}=\delta^\beta_\alpha
\end{equation}
adapted to the geodesics at hand and its dual
coframe (orthonormal as in Eq.~\eqref{Dmet})
\begin{equation}
\label{coframeRanders}
\hat{e}^0 =  \mathrm{d}t-b ,\quad  \hat{e}^\alpha=E^\alpha_{\hphantom{\alpha}i}\mathrm{d}x^i, \quad E^\alpha_{\hphantom{\alpha}i}E^\beta_{\hphantom{\beta}i}\delta_{\alpha\beta}=a_{ij}.
\end{equation}
This will be referred to as the Papapetrou--Randers frame. 

The constant-norm Killing vector field $\check{u}=\check{e}_0=\partial_t$ (with $\hat{u}= -\hat{e}^0 =-\mathrm{d}t+b$) defines a geodesic congruence 
(the orbits of all observers at rest in the Papapetrou--Randers  frame). As was shown in Sec.~\ref{vpf},  
the latter has \emph{zero shear and expansion},
but non-trivial vorticity (see Eqs. (\ref{def24}), (\ref{def3})):
\begin{equation}\label{vortic}
\omega=\frac{1}{2}\mathrm{d} b\quad\Rightarrow\quad 
\omega_{0i}=0,\quad
\omega_{ij}=\frac{1}{2}\left(
\partial_ib_j -\partial_jb_i 
\right).
\end{equation}
The physical effect of vorticity is seen in the obstruction to the parallel transport of the spatial frame $\check{e}_\alpha$ along the congruence:
\begin{equation}
\label{cov_alpha}
\nabla_{\check{e}_0}\check{e}_\alpha=\omega_{\alpha \beta}^{\mathrm{PR}}\delta^{\beta\gamma}\check{e}_\gamma
\quad\Leftrightarrow\quad
\nabla_{\partial_t}\partial_i = \omega_{ij}a^{jk}\left(\partial_k+b_k\partial_t\right)
\end{equation}
($\omega_{ij}$ given in \eqref{vortic} are the spacetime components of the vorticity, while $\omega_{\alpha\beta}^{\mathrm{PR}}=E_\alpha^{\hphantom{\alpha}i} E_\beta^{\hphantom{\beta}j}\omega_{ij}$ are its components in the Papapetrou--Randers frame). Embarked gyroscopes undergo a rotation.

Papapetrou--Randers metrics do not exhibit ergoregions since\footnote{Ergoregions would require a conformal factor in \eqref{Papa} that could vanish and become negative.}  $g_{00}=-1$. However, regions where hyperbolicity is broken (\emph{i.e.}~where constant-$t$ surfaces become timelike) are not excluded. This happens whenever there exist regions where $b_i b_j a^{ij}>1$. Indeed, in these regions, the spatial metric $ a_{ij} - b_i b_j$ possesses a negative eigenvalue, and constant-$t$ surfaces are no longer spacelike.  Therefore the extension of the physical domain accessible to the inertial observers moving along $\check u= \partial_t$ is limited to spacelike disks in which  $b_i b_j a^{ij}<1$ holds. We will come back to this important issue in Secs.~\ref{anal} and \ref{geoflu}.

Before moving to the next topic, we would like to make a last remark. Following Eqs. \eqref{perfeom-dissless}, the shearless and 
expansionless geodesic congruence under consideration could describe the fluid lines of a dissipationless stationary, conformal  
fluid, under the assumption that energy (and pressure) be conserved and constant all over space. As we will see in Sec.~\ref{holog}, 
this is exactly the dynamics that emerges through holography.

\subsection{Zermelo frame}\label{Zer}

Trading the data $(a_{ij},b_i)$ for $(h_{ij},W^i)$ defined as
\begin{eqnarray}
\label{ZR1}
&h_{ij}=\frac{a_{ij}-b_i b_j}{\gamma^2},\quad 
h^{ik}h_{kj}=\delta^i_j,&  \\
&W^i=-\gamma^2b^i , \quad W_i=h_{ij}W^j =-\frac{b_i }{\gamma^2} ,&\label{ZR2}
\end{eqnarray}
the Papapetrou--Randers metric \eqref{Papa} can be recast in the following form
\begin{equation}
\label{Zermelo}
\mathrm{d}s^2=\gamma^2\left[-\mathrm{d}t^2+h_{ij}\left(\mathrm{d}x^i-W^i \mathrm{d}t\right)\left(\mathrm{d}x^j-W^j \mathrm{d}t\right)\right].
\end{equation}
The latter is called Zermelo metric because it first appeared in the framework of the Zermelo problem \cite{Zer31}, yet another member of the relationship network mentioned above\footnote{The Zermelo problem is formulated as follows: find the minimal-time navigation road  on a geometry $\mathrm{d}\ell^2 = h_{ij}\mathrm{d}x^i \mathrm{d}x^j$, in the presence of a moving fluid creating a drift current (wind or tide) $W=W^i\partial_i$ with a ship of fixed propelling velocity (\emph{i.e.}~fixed with respect to the frame comoving with the fluid or, put differently,  of given power).  The answer is reached by searching for null geodesics of \eqref{Zermelo}.\label{Zernav}}.

The Zermelo form of the metric suggests the following orthonormal coframe and its dual frame: 
\begin{eqnarray}
\label{coframeZermelo}
&\hat{z}^0=\gamma {\mathrm{d}}t \, , \quad \hat{z}^\alpha= L^\alpha_{\hphantom{\alpha}i}({\mathrm{d}}x^i - W^i {\mathrm{d}}t), \quad
L_\alpha^{\hphantom{\alpha}i}L^\beta_{\hphantom{\beta}i}=\delta^\alpha_\beta,&\\
\label{frameZermelo}
&\check{z}_0=\frac{1}{\gamma}\left(\partial_t+W^i\partial_i\right)\, , \quad  \check{z}_\alpha=L_\alpha^{\hphantom{\alpha}i}\partial_i, \quad
 L^\alpha_{\hphantom{\alpha}i} L^\beta_{\hphantom{\beta}j}\delta_{\alpha\beta}  =\gamma^2 h_{ij}.&
\end{eqnarray}
We will call the latter the \emph{Zermelo frame}.  Its timelike vector field $\check{z}_0$ defines a congruence of accelerated lines ($\nabla_{\check{z}_0}\check{z}_0\neq 0$). Thus, this frame in not inertial. It is instructive to compare the Papapetrou--Randers frame introduced previously (in \eqref{frameRanders}, \eqref{coframeRanders}) with the Zermelo frame at hand. Being both orthonormal, they are related by a local Lorentz transformation, as one sees by combining the above formulas:
\begin{eqnarray}
\label{RandersZermelot}
\check{e}_0 &=& \gamma\left(\check{z}_0 -W^\beta \check{z}_\beta \right),\\
\check{e}_\alpha &=& \Gamma_\alpha^{\hphantom{\alpha}\beta}
\left(  \check{z}_\beta  -W_\beta \check{z}_0+\frac{\gamma^2-1}{\gamma^2}\left(\frac{W_\beta W^\gamma}{W^2} - \delta_\beta^\gamma\right)\check{z}_\gamma \right),
\label{RandersZermelos}
\end{eqnarray}
where 
\begin{eqnarray}
\label{lorentz}
& \Gamma_\alpha^{\hphantom{\alpha}\beta} =\gamma ^2 E_\alpha^{\hphantom{\alpha}i}L^\beta_{\hphantom{\beta}i}, \quad W^\alpha =\frac{1}{\gamma}L^\alpha_{\hphantom{\alpha}i}W^i, \quad W_\alpha =
 \delta_{\alpha \beta}W^\beta,&\\
 & W^2 = W^\alpha W_\alpha= W^i W_i=1-\frac{1}{\gamma^2}.&
\end{eqnarray}

The interpretation of these expressions is clear. Each spacetime point is the intersection of two lines, belonging each to the two congruences under consideration. At this point $W^\alpha $
are the spatial velocity components of the inertial observer in the spatial frame of the accelerated observer and $\nicefrac{1}{\gamma^{2}}=1-W^2$ the corresponding Lorentz factor.

It is worth making several further comments. The synchronous hypersurface for the inertial observer is by definition dual to the time vector. Since $\mathrm{d}t(\partial_i)=0$ (equivalent to  $\hat{z}^0( \check{z}_\alpha)=0$), this hypersurface is spanned by $\{\partial_i\}$, and hence  is not orthogonal to the inertial congruence 
($\partial_t\cdot \partial_i=b_i$). The orthogonal lines to the Papapetrou--Randers (inertial) observer's synchronous hypersurface are nothing but the accelerated congruence tangent to the vector field $\check{z}_0$ (defining the corresponding Zermelo frame) because $\check{z}_0\cdot \partial_i=0$, whereas the hyperplanes orthogonal to the the inertial congruence (tangent to $ \check{e}_0=\partial_t$) are spanned by  $\{b_i\partial_t+\partial_i\}$. Therefore, since $(\mathrm{d}t-b)(\partial_i+b_i\partial_t)=0$
(equivalent to $\hat{e}^0( \check{e}_\alpha)=0$), the time $\tau$ of Zermelo observers, \emph{i.e.}~the dual of the hypersurface everywhere tangent to the latter hyperplanes, would satisfy  
$\mathrm{d}\tau=\mathrm{d}t-b$. Such a time cannot be defined since $\mathrm{d}b=2\omega\neq 0$. Put differently, no hypersurface exists tangent to the hyperplanes spanned by $\{b_i\partial_t+\partial_i\}$ -- Fr\"obenius theorem. This is a well known manifestation of vorticity.

The last statement again distinguishes the Papapetrou--Randers and Zermelo observers, which are otherwise dual to each other. As for the perception of the rotation, Papapetrou--Randers observers feel it through embarked gyroscopes (see \eqref{cov_alpha}), whereas their inertial motion as witnessed by Zermelo observers satisfies 
\begin{equation}
\label{Zermacc}
\nabla_{\check{z}_0}\check{u}
=\omega_{0\alpha}^{\mathrm{Z}}\delta^{\alpha\beta}\check{z}_ \beta.
\end{equation}
Here $\check{u}=\check{e}_0$ is the velocity of the inertial observers, while  $\omega_{ab}^{\mathrm{Z}}$ are the vorticity components as 
observed in the Zermelo frame: $\omega_{\alpha\beta}^{\mathrm{Z}}=L_\alpha^{\hphantom{\alpha}i}L_\beta^{\hphantom{\beta}j}\omega_{ij}$ and $ \omega_{0\beta}^{\mathrm{Z}}=W^\alpha\omega_{\alpha\beta}^{\mathrm{Z}}$. Hence, for the accelerated observers, the inertial ones are subject to a Coriolis force: Zermelo observers are rotating themselves. The velocity vector $\check{u}=\check{e}_0$ of the inertial observers undergoes a precession around the worldline of a Zermelo observer tangent to $\check{z}_0$. The latter being accelerated, the variation of $\check{u}$ is actually better captured as a Fermi derivative along $\check{z}_0$:
\begin{equation}
\label{Fermi_accel_Zerm}
\mathrm{D}_{\check{z}_0}\check u=\left( \omega_{0\alpha}^{\mathrm{Z}}-\check{z}_\alpha(\gamma)\right)\delta^{\alpha\beta}\check{z}_\beta + W^\alpha \check{z}_\alpha(\gamma)\check{z}_0\, ,
\end{equation}
where $\check{z}_\alpha(\gamma)= L_\alpha^{\hphantom{\alpha}i}\partial_i\gamma$. The extra terms result from the rotation  of the Zermelo frame and contribute to the observed precession of the velocity vector $\check{u}$.  

One can try to tune rotating frames so as to make the perceived angular momentum of a given congruence disappear, 
\emph{i.e.}~make its Fermi derivative vanish with respect to the rotating frame. This leads to the so called zero angular momentum frames  
(ZAMO \cite{Bardeen}).  In general, Zermelo frames \emph{are not} ZAMO frames 
for the Papapetrou--Randers congruences, as the Fermi derivative \eqref{Fermi_accel_Zerm} is generically non-zero. It can however be zero 
under the necessary and sufficient condition
\begin{equation}
\label{ZAMO}
W^j \omega_{ji} = \gamma \partial_i \gamma,
\end{equation}
since this implies that the combination $\omega_{0\alpha}^{\mathrm{Z}}-\check{z}_\alpha(\gamma)$ 
as well as the coefficient of $\check{z}_0$
vanish. Equation  \eqref{ZAMO} carries intrinsic information about the background and can indeed be recast as
\begin{equation}
\label{ZAMO-intrin}
\mathcal{L}_{\check{z}_0}  \hat{e}_0=0.
\end{equation}
When fulfilled, the Zermelo observers coincide with the locally non-rotating (or ZAMO) frames  \cite{Bardeen}. Remarkably, this occurs for a particular case that will be discussed in our subsequent developments.

\subsection{Further properties and analogue gravity interpretation}\label{anal}

In the above analysis and particularly in the change of frame from Papapetrou--Randers to Zermelo, it has been implicitly assumed that $W^2<1$. The velocity of Papapetrou--Randers observers with respect to the Zermelo frame is however dictated by the geometry itself since $W^2=b_i b^i$, and nothing \emph{a priori}  guarantees that  $b_i b^i<1$ everywhere. There are regions in $x$-space where indeed $b_i b^i>1$ bounded by a hypersurface where $b_i b^i=1$. The latter was called \emph{velocity-of-light} hypersurface in \cite{Gibbons} since this is the edge where the Papapetrou--Randers  observer reaches the speed of light with respect to the Zermelo frame. 

The problem raised here is a manifestation of the global hyperbolicity breakdown. Indeed, we have seen that in geometries of the Papapetrou--Randers form \eqref{Papa}, 
constant-$t$ surfaces are not everywhere spacelike. The extension of the physical domain accessible to the inertial observers moving along $\check u= \partial_t$ is limited to spacelike disks in which  $b^2<1$ holds, bounded by the velocity-of-light surface, where these observers become luminal.

The breaking of hyperbolicity is usually accompanied with the appearance of closed timelike curves (CTCs). These are ordinary spacelike circles, lying in constant-$t$ surfaces, which become timelike when these surfaces cease being spacelike, \emph{i.e.}~when $b^2>1$. The CTCs at hand differ in nature from those due to compact time (as in the $SL(2,\mathbb{R})$  group manifold), and cannot be removed by unwrapping time. They require an excision procedure for consistently removing, if possible, the $b^2>1$ domain, in order to keep a causally safe spacetime. This is comparable to what happens in the case of the three-dimensional Ba\~nados--Teitelboim--Zanelli black hole \cite{Banados:1992wn} -- although in the latter case the trouble is not due to hyperbolicity issues. We will come back to the CTCs when studying the anti-de Sitter Taub--NUT geometry.

Although the issue of hyperbolicity is intrinsic to our stationary geometries, moving from the form \eqref{Papa} to the form \eqref{Zermelo} may provide alternative or complementary views. In the Zermelo form \eqref{Zermelo} the trouble is basically encapsulated in the conformal factor. However, some problems such as the original Zermelo navigation problem referred to in footnote~\ref{Zernav} are sensitive to the general conformal class\footnote{In the Zermelo navigation problem we look for null geodesics. In that framework, going to regions where $\gamma^2<0$ means having a drift current faster than what the ship can overcome.} of \eqref{Zermelo}, and the conformal factor  $\gamma^2$ can be dropped or replaced. Doing so can  leave us with a geometry potentially sensible everywhere. This instance appears precisely in analogue gravity systems. 

Metrics of the form  \eqref{Zermelo} are in fact known as acoustic or optical (see the original works \cite{Unruh:1980cg, Unruh:1994je} or e.g.~\cite{Barcelo} for an up-to-date review). They are used for describing the propagation of sound/light disturbances in relativistic or non-relativistic fluids moving with velocity $W^i$ in spatial geometry $h_{ij}$, and subject to appropriate thermodynamic/hydrodynamic assumptions. In this approach, the full metric  (\ref{Zermelo}) is an \emph{analogue} metric and is \emph{not} the actual metric of physical spacetime.
Under this perspective, peculiarities such as CTCs, potentially present in the analogue geometry, have no real, physical existence. They are manifestations of other underlying physical properties such as supersonic/superluminal regimes in the flowing medium.

In order to be concrete, we would like to quote two examples. The first is the original one, where a fluid is flowing on a geometry  $\mathrm{d}\ell^2 = h_{ij}\mathrm{d}x^i \mathrm{d}x^j$ with velocity field $W=W^i\partial_i$. Assuming the fluid is non-relativistic, isolated, with mass density 
 $\varrho$ and pressure $p$, in barotropic evolution (\emph{i.e.}~such that the enthalpy variations satisfy $\mathrm{d}h=\nicefrac{\mathrm{d}p}{\varrho}$) and with sound velocity $c_{\mathrm{s}}=\nicefrac{1}{\sqrt{\nicefrac{\partial \varrho}{\partial p}}}$, one finds that irrotational acoustic perturbations propagate along null geodesics of the metric 
\begin{equation}\label{Zermacous}
 \mathrm{d}s^2= \frac{\varrho}{c_{\mathrm{s}}} \left(-c_{\mathrm{s}}^2\mathrm{d}t^2+h_{ij}
\left(\mathrm{d}x^i-W^i \mathrm{d}t \right)\left(\mathrm{d}x^j-W^j \mathrm{d}t \right)\right),
\end{equation}
and satisfy the corresponding scalar field equation. The metric \eqref{Zermacous} is of the form \eqref{Zermelo}. It is however analogue and not the actual spacetime geometry, which is  Galilean.  This analogue metric is called acoustic in the case at hand. Similar results are available for light propagation, leading to optical geometries (see e.g.~\cite{Liberati, Cacciatori}).

Similarly we can quote the case of a relativistic conformal fluid, at rest in the Papapetrou--Randers frame of a Papapetrou--Randers geometry \eqref{Papa} -- with $B=1$. In this case, fluid lines are tangent to $\check{u}=\partial_t$, the kinematics is shearless and expansionless with vorticity \eqref{vortic}, and thus it is non-dissipative. It is also geodesic (see \eqref{accRan}) and Eqs. \eqref{perfeom-dissless} imply that $\varepsilon, p$ and $T$ are constant everywhere. The propagation of irrotational perturbations in this set up is captured by the following acoustic metric\footnote{This result is easy to establish, following e.g.~similar reasoning as in \cite{Shapere}.}:
\begin{equation}
\label{Papacou}
\mathrm{d}s^2=\frac{T^{D-2}}{\sqrt{D-1}}\left(-(\mathrm{d}t-b_i \mathrm{d}x^i)^2+(D-1) a_{ij}(x)\mathrm{d}x^i \mathrm{d}x^j\right),
\end{equation}
of the Papapetrou--Randers form, which can be recast in the Zermelo form \eqref{Zermelo}, following \eqref{ZR1} and \eqref{ZR2}.

\section{Examples in $\mathbf{2+1}$ dimensions}\label{ex21}

Examples of Papapetrou--Randers geometries are numerous, possessing diverse properties regarding their isometries, their curvature, the regularity of their Randers--Zermelo transformation, etc. Their expressions can be simple in Papapetrou--Randers form and complicated in the Zermelo representation or vice versa. 
All this depends in particular on the dimension\footnote{See \cite{Gibbons} for a detailed account of properties and examples in $3+1$ dimensions.}. We will here focus on a few three-dimensional examples that turn out to emerge as holographic duals of exact four-dimensional bulk spacetimes.
Although the nature of the boundary three-dimensional spacetime regarding Einstein's equations plays little role in holography, some underlying intrinsic properties appear to be generic for the backgrounds under consideration, and would deserve further investigation.
Furthermore, all examples below are homogeneous spaces\footnote{Following \cite{Milnor,Scott}, homogeneous  three-manifolds include all 9
Bianchi groups plus 3 coset spaces, which are $H_3$, $H_2\times S^1$,
$S^2\times S^1$ ($S^n$ and $H_n$ are spheres and hyperbolic spaces respectively).}, even though neither was homogeneity an
\emph{a priori} criterion, nor did our list provide an exhaustive classification of stationary backgrounds with a spatially homogeneous timelike Killing vector field.

\subsection{Warped three-spheres: Bianchi IX}\label{S3}

Warped three-spheres, earlier more appropriately called biaxially squashed three-spheres, are deformations of the standard homogeneous and isotropic (round) $S^3$. This deformation breaks the original $SU(2)\times SU(2)$ isometry group down to $SU(2)$ or $SU(2)\times U(1)$. These spaces can be endowed with the metric 
\begin{equation}
\label{homfol}
\mathrm{d}s^2 =  
    \sum_{i=1}^{3} \left(\gamma_i\sigma^i\right)^2, 
\end{equation}
where  $\gamma_i $ are constants and  $\sigma^i $ are the left-invariant Maurer--Cartan forms of $SU(2)$. In terms of Euler angles $0\leq\theta\leq \pi, 0\leq\phi\leq 2\pi, 0\leq\psi\leq
{4\pi}$, these one-forms read:
\begin{equation}
  \begin{cases}
  \label{LMC}
\sigma^1= \sin\theta \sin\psi \, \mathrm{d}\phi+\cos \psi \, \mathrm{d}\theta \\
\sigma^2= \sin\theta\cos\psi\, \mathrm{d}\phi-\sin\psi\, \mathrm{d}\theta\\
\sigma^3=\cos\theta\, \mathrm{d}\phi+\mathrm{d}\psi.
\end{cases}
\end{equation}

For reasons that will become clear in the following, we will consider situations where $\gamma_1=\gamma_2$. The spaces obtained in this way are called Berger spheres. They are axisymmetric \emph{i.e.}~have  $SU(2)\times U(1)$ isometry group. Since we  are interested 
in spaces with Lorentzian signature, we must  set negative $\gamma_3^2$ and the metric finally reads:
\begin{equation}
\mathrm{d}s^2 =
 L^2\left[(\sigma^1)^2+
\left(\sigma^2\right)^2\right]-4n^2\left(\sigma^3\right)^2
=-4n^2\left( \mathrm{d}\psi + \cos \theta  \mathrm{d}\phi\right)^2 + L^2\left(
 \mathrm{d}\theta^2 +
\sin^2\theta \mathrm{d}\phi^2
\right),\label{berg-lor-eul}
\end{equation}
where $2L$ is the radius of the original undeformed $S^3$ and $2nk$ the deformation parameter ($k=\nicefrac{1}{L}$). 

The time coordinate in \eqref{berg-lor-eul} is $\psi$. In the original Euclidean sphere this was an angle, but there is no reason to keep it compact in the Lorentzian version at hand\footnote{This point deserves some comments, which we postpone to the discussion on the bulk geometries leading to \eqref{berg-lor-eul} as boundary.}. Introducing a non-compact time $t=-2n (\psi+\phi)$, the metric \eqref{berg-lor-eul} assumes the form:
\begin{equation}
\label{berg-lor-t}
\mathrm{d}s^2 = -\left( \mathrm{d}t + 4n\sin^2 \nicefrac{\theta}{2}\,  \mathrm{d}\phi\right)^2+ L^2 \left(
 \mathrm{d}\theta^2 +
\sin^2\theta \mathrm{d}\phi^2
\right).
\end{equation}
This metric is of Papapetrou--Randers type \eqref{Papa}. The base $\mathrm{d}x^2=a_{ij}(x)\mathrm{d}x^i \mathrm{d}x^j$ is a two-sphere of radius $L$, while $b= -4n\sin^2 \nicefrac{\theta}{2}\,\mathrm{d}\phi$ is a Dirac-monopole-like potential. The latter creates a constant -- in the Papapetrou--Randers orthonormal coframe -- vorticity 
\begin{equation}
\label{TNvort}
\omega = -n\sin\theta\mathrm{d}\theta\wedge
 \mathrm{d}\phi
 = -nk^2\,\hat{e}^1\wedge \hat{e}^2 
\end{equation}
for the geodesic congruence tangent to $\check{u}\equiv\check{e}_0=\partial_t$ (up to a delta-function contribution of the ``Misner point'' at the southern pole -- this name will be justified later).

As already quoted, the space at hand is homogeneous and belongs to the family of spaces invariant under a four-parameter group of motions \cite{rayPRD80, rebPRD83}, here generated by the following Killing vectors:
 \begin{equation}
\begin{cases}
  \label{LKilR2}
\xi_1= - \sin\phi \cot\theta\, \partial_\phi+\cos \phi\, \partial_\theta-2 n \frac{\sin \phi}{\sin \theta}(1-\cos \theta) \, \partial_t \\
\xi_2=  \cos\phi \cot\theta\, \partial_\phi+\sin \phi \,\partial_\theta+2 n \frac{\cos \phi}{\sin \theta}(1-\cos \theta) \, \partial_t  \\
\xi_3= \partial_\phi-2 n \, \partial_t\\
e_3= -2 n \, \partial_t.
\end{cases}
\end{equation}
The three former generate the left $SU(2)$, whereas the latter generates an extra $\mathbb{R}$ factor (instead of $U(1)$ since $t$ is non-compact). 

The background \eqref{berg-lor-t} is not globally hyperbolic. Even though it is homogeneous, constant-$t$ surfaces are not and  $\gamma= \nicefrac{1}{\sqrt{1-4n^2k^2\tan^2\nicefrac{\theta}{2}}}$ diverges 
when $\theta$ reaches $\theta_*=2\arctan \nicefrac{L}{2n}$. Hyperbolicity holds in the disk $0<\theta<\theta_*$, whereas it breaks down in the complementary disk ($\theta_*<\theta<\pi$) centered at the Misner point, where $\partial_\phi$ becomes timelike. As a consequence, the circles tangent to $\partial_\phi$ become CTCs for $\theta_*<\theta<\pi$. 
Homogeneity implies furthermore that CTCs are present everywhere, passing through any arbitrary point of spacetime. In 
particular, for  $0<\theta<\theta_*$ the CTCs are sections of cylinders normal to the constant-$t$ surfaces. The time coordinate $t$ evolves periodically along these elliptically shaped CTCs. 

As we will see in the forthcoming sections,  the situation described here is quite generic for three-dimensional homogeneous spacetimes. These include 
the case of Som--Raychaud- huri (Bianchi II) and the celebrated G\"odel space (Bianchi VIII). They are illustrative examples of how homogeneity combined with rotation often leads to the breakdown of hyperbolicity and the emergence of CTCs.
G\"odel space in particular was the first to be recognized as plagued by CTCs. The CTCs present in these spaces, however, are not geodesics \cite{rebPRD83, Gibbons:1999uv, Drukker:2003mg}. Their presence is therefore harmless for classical causality. This is why G\"odel-like solutions like the case under consideration have never been truly discarded, leaving open the possibility of quantum mechanical validity\footnote{Attempts, among others in string theory within holography, were proposed a few years ago (see e.g.~\cite{Drukker:2003mg, Hikida:2003yd, Israel:2003cx,Israel:2004vv, Israel:2004cd,Caldarelli:2004mz} and references therein).}.

Let us incidentally mention that the above three-dimensional geometry \eqref{berg-lor-t} -- when lifted to four dimensions by 
taking the direct product with an extra flat direction -- has been shown to satisfy Einstein's equations with cosmological 
constant and energy--momentum tensor produced by some specific charged fluid \cite{rebPRD83}. Alternatively, it also 
satisfies the equations of topologically massive gravity  \cite{cs}, subject to a Kerr--Schild \cite{KS65} deformation\footnote{It should be mentioned that the metric \eqref{berg-lor-t} solves also topological massive gravity equations without Kerr--Schild deformation~\cite{Banados:2005da,Anninos:2008fx,Anninos:2011vd}, provided one tunes appropriately the relationship among $L^2$ and $k^2$. This holds for generic  warped homogeneous 
spaces like $\mathrm{AdS}_3$, studied in Sec. \ref{ads3}.}
 \begin{equation}
  \label{TMG}
R_{\mu\nu} -\frac{R}{2} g_{\mu\nu} +\Lambda^{(3)} g_{\mu\nu}= \frac{1}{4nk^2} C_{\mu\nu}+\frac{k^2\left(1+4n^2k^2\right)}{4}u_\mu u_\nu,
\end{equation}
where $\hat{u}=-\hat{e}_0=-\mathrm{d}t - 4n\sin^2 \nicefrac{\theta}{2}  \mathrm{d}\phi$, $\Lambda^{(3)} = \nicefrac{k^2}{4} $ is the cosmological constant for an undeformed $S^3$ and $C_{\mu\nu}$ are the components of the Cotton--York tensor defined as
 \begin{equation}
 C^{\mu\nu}=\frac{\epsilon^{\mu\rho\sigma}}{\sqrt{-g}}
\nabla_\rho\left(R^{\nu}_{\hphantom{\nu}\sigma}-\frac{1}{4}R\delta^{\nu}_\sigma \right) .
\end{equation}
For the background \eqref{berg-lor-t} this tensor is  
 \begin{equation}\label{CotSU}
 C_{\mu\nu}\mathrm{d}x^\mu\mathrm{d}x^\nu = nk^4\left(1+4n^2k^2\right)\left[2\hat{u}^2
 + L^2 \left(
 \mathrm{d}\theta^2 +
\sin^2\theta \mathrm{d}\phi^2 
 \right)  \right].
\end{equation}
For later convenience we also quote:
 \begin{equation}\label{RicSU}
R_{\mu\nu}\mathrm{d}x^\mu\mathrm{d}x^\nu  = 2k^4 n^2 \hat{u}^2+
\left(1+2n^2k^2\right) \left(
 \mathrm{d}\theta^2 +
\sin^2\theta \mathrm{d}\phi^2 
 \right)
\end{equation}
and 
 \begin{equation}\label{RSU}
R= 2k^2
\left(1+n^2k^2\right). 
\end{equation}

\subsection{Warped AdS$_3$: Bianchi VIII} \label{ads3}

Following the paradigm of squashed three-spheres, studied in Sec.~\ref{S3}, we analyze here the deformations of $\mathrm{AdS}_3$. The latter is the (universal covering of the) $SL(2,\mathbb{R})$
group manifold and has left and right $SL(2,\mathbb{R})$ isometries. Homogeneous deformations can break partially or completely one $SL(2,\mathbb{R})$ factor. 

There are many realizations of the Maurer--Cartan forms of $SL(2,\mathbb{R})$. The one we choose here  is convenient for the specific deformation we will consider in the following:
\begin{equation}
\begin{cases} \label{LMCSL}
\rho^0=-\mathrm{d}\tau+\cosh\sigma\, \mathrm{d}\phi\\
\rho^1=  -\sin\tau\, \mathrm{d}\sigma- \sinh\sigma\cos\tau\, \mathrm{d}\phi\\
\rho^2=\cos \tau \, \mathrm{d}\sigma + \sinh\sigma \sin\tau \, \mathrm{d}\phi,
\end{cases}
\end{equation}
where $0\leq\phi\leq 2\pi, 0\leq\sigma< +\infty$, and $\tau \in [0,2\pi]$ or better $\mathbb{R}$ if we consider the universal covering of the space. The metrics under consideration are of the form 
 \begin{equation}
\label{homfolSL}
\mathrm{d}s^2 =
   - \left(\gamma_0\rho^0\right)^2
   + \left(\gamma_1\rho^1\right)^2+ \left(\gamma_2\rho^2\right)^2 .
\end{equation}
The $\gamma$s being constant,  this geometry is homogeneous. When $\forall i, \gamma_i=L$, we recover radius-$2L$ $\mathrm{AdS}_3$. 

We will restrict ourselves here to the elliptically-squashed  $\mathrm{AdS}_3$, obtained with $\gamma_0=2p\in \mathbb{R}$ and $\gamma_1=\gamma_2=L$. These geometries have a 4-parameter isometry group  $SL(2,\mathbb{R})\times \mathbb{R}$. The Abelian factor is the remaining one-parameter subgroup of the broken isotropy symmetry --  $U(1)$ before taking  the universal covering. 

Elliptic deformations were studied as such in \cite{Rooman:1998xf} and the continuous line obtained there coincides -- when uplifted to four dimensions -- with the family of spacetime-homogeneous G\"odel-type metrics discussed in \cite{rebPRD83}.
After a coordinate transformation trading $\tau$ for $t=2p(\tau-\phi)\in \mathbb{R}$, the metric  \eqref{homfolSL} reads:
\begin{equation}
\label{SLell}
\mathrm{d}s^2 = -\left( \mathrm{d}t - 4p\sinh^2 \nicefrac{\sigma}{2}\,  \mathrm{d}\phi\right)^2+ L^2 \left(
 \mathrm{d}\sigma^2 +
\sinh^2\sigma\, \mathrm{d}\phi^2
\right).
\end{equation}
This is a timelike fibration over a hyperbolic plane $H_2$.

Alternative inequivalent deformations exist in $SL(2,\mathbb{R})$, such as the hyperbolic or the parabolic ones which lead to spacelike fibrations over $\mathrm{AdS}_2$ or plane-wave superpositions with  $\mathrm{AdS}_3$. They have been studied extensively in the framework of string theory \cite{Israel:2003cx, Israel:2004vv,Israel:2004cd} and more recently in holography \cite{Anninos:2011vd}.   However, these are not of the Papapetrou--Randers form \eqref{Papa}, as opposed to the elliptic deformation for which $b= 4p\sinh^2 \nicefrac{\sigma}{2}\,  \mathrm{d}\phi$ while the base
$\mathrm{d}x^2=a_{ij}(x)\mathrm{d}x^i \mathrm{d}x^j$ is radius-$L$ Lobatchevsky plane. The ``non-compact Dirac potential'' $b$  creates  for the geodesic congruence tangent to $\check{u}\equiv\check{e}_0=\partial_t$ a homogeneous vorticity, which -- in the Papapetrou--Randers orthonormal coframe -- reads:
\begin{equation}
\label{hypvort}
\omega = p\sinh\sigma\,\mathrm{d}\sigma\wedge
 \mathrm{d}\phi
 = pk^2\,\hat{e}^1\wedge \hat{e}^2.
\end{equation}
In the case at hand, the base is non-compact and the Misner point is rejected to infinity.  

The hyperbolicity properties of the background \eqref{SLell} are richer than for \eqref{berg-lor-t}. Indeed, for the space at hand, the Lorentz factor 
relating Zermelo and Papapetrou--Randers observers is 
$\gamma= \nicefrac{1}{\sqrt{1-4p^2k^2\tanh^2\nicefrac{\sigma}{2}}}$. This factor remains finite for finite $\sigma$ provided $p\leq \nicefrac{L}{2}$ -- 
the limiting case $p = \nicefrac{L}{2}$ corresponding to the undeformed $\mathrm{AdS}_3$. Under this condition, the space is globally 
hyperbolic and constant-$t$ surfaces are spacelike everywhere. When  $p > \nicefrac{L}{2}$ this property breaks down for 
$\sigma>\sigma_*=2\mathrm{arctanh}\,  \nicefrac{L}{2p}$. Constant-$t$ surfaces are spacelike on a disk only, centered at $\sigma=0$ and bounded by a velocity-of-light surface located at $\sigma=\sigma_*$. The breakdown of hyperbolicity is accompanied with the appearance 
of CTCs, present everywhere as a consequence of homogeneity.  This happens in particular for $p = \nicefrac{L}{\sqrt{2}}$ corresponding to G\"odel's solution. 
 
As mentioned earlier, the absence of hyperbolicity in the backgrounds considered in our general framework is closely related to the 
combination of vorticity and homogeneity. This is even better illustrated in the $\mathrm{AdS}_3$, where the non-compact nature of 
the base makes it possible to evade the breakdown, provided the vorticity is small enough with respect to the scale $L$ set by the
 curvature. The isometry group is in the present case $SL(2,\mathbb{R})\times \mathbb{R}$ 
generated by the following Killing vectors:
 \begin{equation}
\begin{cases}
  \label{LKilSL}
\zeta_0= \partial_\phi-2 p \, \partial_t\\
\zeta_1= -\cos\phi \coth\sigma\, \partial_\phi-\sin \phi \,\partial_\sigma-2 p \frac{\cos \phi}{\sinh \sigma}(1-\cosh \sigma) \, \partial_t  \\
\zeta_2=  - \sin\phi \coth\sigma\, \partial_\phi+\cos \phi\, \partial_\sigma-2 p \frac{\sin \phi}{\sinh \sigma}(1-\cosh \sigma) \, \partial_t \\
q_0= -2 p \, \partial_t.
\end{cases}
\end{equation}
For completeness we also quote the Levi--Civita curvature properties of the metric \eqref{SLell}:
 \begin{eqnarray}\label{CotSL}
C_{\mu\nu}\,\mathrm{d}x^\mu\mathrm{d}x^\nu &=& pk^4\left(1-4p^2k^2\right)\left[2\hat{u}^2
 + L^2 \left(
 \mathrm{d}\sigma^2 +
\sinh^2\sigma\, \mathrm{d}\phi^2 
 \right)  \right],\\
 \label{RicSL}
R_{\mu\nu}\,\mathrm{d}x^\mu\mathrm{d}x^\nu  &=& 2k^4 p^2 \hat{u}^2-
\left(1-2p^2k^2\right) \left(
 \mathrm{d}\sigma^2 +
\sinh^2\sigma\, \mathrm{d}\phi^2 
 \right),\\
\label{RSL}
R&=& -2k^2
\left(1-p^2k^2\right),
\end{eqnarray}
where $\hat{u}=-\mathrm{d}t+4p\sinh^2 \nicefrac{\sigma}{2}\,  \mathrm{d}\phi$.

\subsection{Rotating Einstein universes}

In the limit $n \to 0 $, the geometry \eqref{berg-lor-t} becomes a homogeneous metric on 
 $\mathbb{R}\times S^2$. This is the Einstein static universe, which is a trivial case of static Papapetrou--Randers geometry. The isometry group remains unaltered and generated by the Killing vectors  \eqref{LKilR2}. The metric reads: 
\begin{equation}
\label{ESU}
\mathrm{d}s^2=
-\mathrm{d}t^2+ L^2\left(\mathrm{d}\theta^2+   \sin^2\theta\, \mathrm{d}\phi ^2 \right).
\end{equation}
Trading $(\theta,\phi)$ for $(\theta',\phi')$ defined as
\begin{equation}
\begin{cases}
\phi=\phi'+ \Omega_{\infty}t, \quad
D_{\theta}  \Delta_{\theta'} =\Xi,\quad  \Xi= 1 - L^2 \Omega_{\infty}^2 , \\ D_\theta=1 - L^2 \Omega_{\infty}^2 \sin^2\theta ,\quad
\Delta_{\theta'} =  1 - L^2 \Omega_{\infty}^2 \cos^2\theta', 
\end{cases}
\end{equation}
we obtain
\begin{equation}
\label{ERU-zer}
\mathrm{d}s^2=-\mathrm{d}t^2+ \frac{\Xi L^2}{\Delta_{\theta'}^2}\left(\mathrm{d}\theta'^2+  \frac{\Delta_{\theta'}}{\Xi}  \sin^2\theta' \left[ \mathrm{d}\phi' + \Omega_{\infty} \mathrm{d}t \right]^2 \right).
\end{equation}
This is the Einstein universe uniformly rotating at angular velocity $ \Omega_{\infty}$ in spheroidal coordinates. 

The metric \eqref{ERU-zer} is in a Zermelo form \eqref{Zermelo}. It  can be brought to Papapetrou--Randers form using the general  transformations \eqref{ZR1} and  \eqref{ZR2}:
\begin{equation}
\label{ERU-ran}
\mathrm{d}s^2= \frac{\Xi}{\Delta_{\theta'}}\left(-
      \left[ \mathrm{d}t - \frac{L^2 \Omega_{\infty}}{\Xi} \sin^2\theta'  \mathrm{d}\phi'\right]^2
      + \frac{L^2\mathrm{d}\theta'^2}{\Delta_{\theta'}}
  +  \frac{L^2\Delta_{\theta'}}{\Xi^2} \sin^2\theta' \mathrm{d}\phi'^2\right).
\end{equation}
In order to reach a canonical Papapetrou--Randers form \eqref{Papa} with $B=1$, a conformal transformation is required with conformal factor 
$\Phi= \nicefrac{\Delta_{\theta'}}{\Xi}$. The resulting geometry,
\begin{equation}
\label{KADS-bound}
\mathrm{d}s^2=-
      \left[ \mathrm{d}t - \frac{L^2 \Omega_{\infty}}{\Xi} \sin^2\theta'  \mathrm{d}\phi'\right]^2
      + \frac{L^2\mathrm{d}\theta'^2}{\Delta_{\theta'}}
  +  \frac{L^2\Delta_{\theta'}}{\Xi^2} \sin^2\theta' \mathrm{d}\phi'^2,
\end{equation}
appears as the boundary of Kerr--AdS four-dimensional bulk geometry \cite{Hawking:1998kw}.
We would like to stress that although \eqref{ERU-ran} is still invariant under $SU(2)\times \mathbb{R}$ generated by \eqref{LKilR2} (at $n=0$), this is no longer true for   \eqref{KADS-bound}, where the isometry group is reduced to $U(1)\times \mathbb{R}$ generated by $\partial_\phi$ and $\partial_t$, while  $\xi_1$ and $\xi_2$ in \eqref{LKilR2} become conformal Killing vectors.

Let us finally mention that we can in a similar fashion consider the limit $p\to 0$ in the elliptically 
deformed $\mathrm{AdS}_3$ geometries, given in \eqref{SLell}. This leads to a Einstein-static-universe-like geometry of the type $\mathbb{R}\times H_2$, which can be brought into the form 
\begin{equation}
\label{ERUhyp-ran}
\mathrm{d}s^2= \frac{Z}{\Theta_{\sigma'}}\left(-
      \left[ \mathrm{d}t - \frac{L^2 \Omega_{\infty}}{Z} \sinh^2\sigma'  \mathrm{d}\phi'\right]^2
      + \frac{L^2\mathrm{d}\sigma'^2}{\Theta_{\sigma'}}
  +  \frac{L^2\Theta_{\sigma'}}{Z^2} \sinh^2\sigma' \mathrm{d}\phi'^2\right),
\end{equation}
after a coordinate transformation $(\sigma,\phi)\mapsto(\sigma',\phi')$ with 
\begin{equation}
\begin{cases}
\phi=\phi'+ \Omega_{\infty}t, \quad
H_{\sigma}  \Theta_{\sigma'} =Z,\quad  Z= 1 + L^2 \Omega_{\infty}^2 , \\
H_{\sigma} =  1 - L^2 \Omega_{\infty}^2 \sinh^2\sigma, \quad
\Theta_{\sigma'} =  1 + L^2 \Omega_{\infty}^2 \cosh^2\sigma'.
\end{cases}
\end{equation}
One can similarly perform a conformal transformation leading to 
\begin{equation}
\label{ERUhyp-bound}
\mathrm{d}s^2= -
      \left[ \mathrm{d}t - \frac{L^2 \Omega_{\infty}}{Z} \sinh^2\sigma'  \mathrm{d}\phi'\right]^2
      + \frac{L^2\mathrm{d}\sigma'^2}{\Theta_{\sigma'}}
  +  \frac{L^2\Theta_{\sigma'}}{Z^2} \sinh^2\sigma' \mathrm{d}\phi'^2.
\end{equation}

The geometry described by \eqref{ERUhyp-bound} is globally hyperbolic without any restriction. For the metric \eqref{KADS-bound}, global hyperbolicity holds  under the condition $\Omega_\infty<k$. Again this is a  bound on the magnitude of the vorticity 
\begin{equation}
\label{Kerrvort}
\omega=\frac{L^2\Omega_{\infty}\sin2\theta'}{2\Xi}\mathrm{d}\theta'\wedge \mathrm{d}\phi'
\end{equation}
carried by the geodesic congruence $\check{u}=\check{e}_0=\partial_t$ in this background -- and is relaxed for the non-compact base 
(squashed $H_2$ in \eqref{ERUhyp-bound} versus squashed $S^2$ in \eqref{KADS-bound}), where 
\begin{equation}
\label{HRUhypvort}
\omega=\frac{L^2\Omega_{\infty}\sinh2\sigma'}{2Z}\mathrm{d}\sigma'\wedge \mathrm{d}\phi'.
\end{equation}

\subsection{Zooming at the poles: Bianchi II and VII$_0$}

We can further analyze the geometries met so far  by zooming around their poles. This exhibits new backgrounds of Papapetrou--Randers type, interesting for their own right.
\paragraph{Som--Raychaudhuri and Heisenberg algebra.}
The geometry around $\theta \approx 0$ in \eqref{berg-lor-t} or 
$\sigma \approx 0$ in \eqref{SLell} 
is
\begin{equation}
\label{KAdS-bry-pol}
\mathrm{d}s^2\approx-\left(
\mathrm{d}t - L^2\Omega_{\infty}\chi^2\mathrm{d}\phi
\right)^2 +L^2\left(
\mathrm{d}\chi^2
+\chi^2
\mathrm{d}\phi^2\right)
\end{equation}
with $\chi=\theta$ or $\sigma$, and $\Omega_{\infty}=-nk^2$ or $pk^2$. 
This very same geometry appears also around  $\theta \approx 0$ or $\pi$ in \eqref{KADS-bound} and  around  $\sigma \approx 0$  in \eqref{ERUhyp-bound}  --
upon appropriate definitions of $\chi$ involving $\Xi$ or $Z$.

Metric (\ref{KAdS-bry-pol}) is the Som--Raychaudhuri space, 
found in \cite{SR68} by solving Einstein equations with rotating, charged dust with zero Lorentz force. It belongs to the general family of three-dimensional homogeneous spaces possessing 4 isometries studied in \cite{rayPRD80, rebPRD83}, which include the various metrics that we have discussed so far here.
 In the case of Som--Raychaudhuri (Eq.~(\ref{KAdS-bry-pol})) the isometries are generated by the following Killing vectors:
\begin{equation}
  \begin{cases}
  \label{LKilRn}
K_x=  k\frac{\sin\phi}{\chi} \, \partial_\phi-k\cos \phi\, \partial_\chi- L\Omega_{\infty}\chi\sin \phi  \, \partial_t \\
K_y=  k\frac{\cos\phi}{\chi} \, \partial_\phi+k\sin \phi \,\partial_\chi-L \Omega_{\infty}\chi\cos \phi\, \partial_t  \\
K_0=2 \Omega_{\infty}\, \partial_t\\
K=\partial_\phi.
\end{cases}
\end{equation}
The vectors $K_x, K_y$ and $K_0$ form a Heisenberg algebra, and indeed the Som--Raychaudhuri  metric can be built as the group manifold of the Heisenberg group (Bianchi II) at an extended-symmetry (isotropy) point with an extra symmetry generator\footnote{$\left[K_x, K_y\right]=K_0$, $\left[K_x, K_0\right]=\left[K_y, K_0\right]=0$,  $\left[K, K_x\right]=K_y$, $\left[K, K_y\right]=-K_x$ and $\left[K, K_0\right]=0$.} $\partial_\phi$. Actually, this corresponds to a contraction of $SU(2)\times \mathbb{R}$ into a semi-direct product of the Heisenberg group with an extra $U(1)$ generated by $K_x=-k\xi_1, K_y=k\xi_2, K_0=k^2 e_3, K=\xi_3-e_3$ (see (\ref{LKilR2})). 
It can similarly emerge from the $SL(2,\mathbb{R})\times \mathbb{R}$ algebra \eqref{LKilSL}.

Similarly to G\"odel space, Som--Raychaudhuri  space contains non-geodesic closed timelike curves. These are circles of radius $\chi$ larger than $\nicefrac{1}{\Omega_{\infty}}$ \cite{rebPLA87}. 

\paragraph{Flat vortex and $\mathbf{E(2)}$ geometry.}

The southern pole of \eqref{berg-lor-t} is not captured in the above scheme. Indeed this is a fixed point of the transformation generated by the Killing vector $\xi_3+e_3$ (see \eqref{LKilR2}), which  we referred to as the Misner point. Around this point 
  \begin{equation}
\mathrm{d}s^2\approx -\left(\mathrm{d}t + n\left(4-\chi^2\right) \mathrm{d}\phi\right)^2 +L^2\left( \mathrm{d}\chi^2
+\chi^2
\mathrm{d}\phi^2\right), 
\label{AKT-south}
\end{equation}
where $\chi=\pi-\theta$. The latter is known as a flat vortex geometry, homogeneous and invariant under an $E(2)\times \mathbb{R}$ algebra ($E(2)$ is Bianchi $\mathrm{VII}_0$) generated by\footnote{$\left[L_x,L_y\right]=0$, $\left[L_0,L_x\right]=L_y$,  $\left[L_0,L_y\right]=-L_x$ and everything commutes with $L$.} 
  \begin{equation}
\begin{cases}
  \label{LLilRs}
L_x=k \xi_1= k \frac{\sin\phi}{\chi} \left(\partial_\phi -4n\, \partial_t\right)-k\cos \phi\, \partial_\chi \\
L_y=-k \xi_2=k  \frac{\cos\phi}{\chi} \left(\partial_\phi -4n\, \partial_t\right)+k\sin \phi \,\partial_\chi\\
L_0=\xi_3= \partial_\phi -2n\, \partial_t\\
L= e_3= -2n\, \partial_t.
\end{cases}
\end{equation}
It also appears as a contraction of the $SU(2)\times\mathbb{R}$.

\section{Holography in the Fefferman--Graham expansion: from $\mathbf{3+1}$ to $\mathbf{2+1}$}\label{holog} 

In this section we study the properties of holographic fluids, as they arise in the Fefferman--Graham expansion along the holographic radial coordinate. The set of data reached in this manner contains the boundary frame \emph{i.e.}~the geometrical background hosting the fluid, as well as the energy--momentum tensor, which describes the fluid dynamics. This general method is applied to selected four-dimensional solutions of Einstein's equations, whose boundaries coincide with Papapetrou--Randers geometries studied in Sec.~\ref{ex21}.

\subsection{Split formalism and Fefferman--Graham in a nutshell}

We find  illuminating  to discuss holographic fluid dynamics in $D=2+1$ dimensions starting from the $3+1$-split formalism introduced in \cite{Leigh:2007wf, Mansi:2008br, Mansi:2008bs}. We begin with the Einstein--Hilbert action in the Palatini first-order formulation
\begin{eqnarray}
S&=&-\frac{1}{32\pi G_\mathrm{N}}\int \epsilon_{ABCD}\left(R^{AB}-\frac{\Lambda}{6}E^A\wedge E^B\right)\wedge E^C\wedge E^D\nonumber\\
&=& \frac{1}{16\pi G_\mathrm{N}}\int \mathrm{d}^4x\sqrt{-g}(R-2\Lambda),\label{HPaction}
\end{eqnarray}
where $G_\mathrm{N}$ is Newton's constant. We also assume negative cosmological constant expressed as $\Lambda =-\nicefrac{3}{L^2}=-3k^2$. 
We denote the orthonormal coframe $E^A$, $A=r,a$ and use for the bulk metric the signature $+-++$. The first direction  $r$ is the holographic one and 
$\mathrm{x}\equiv(t,x^1,x^2)\equiv(t,x)$.

Bulk solutions are taken in the Fefferman--Graham form  
\begin{equation}
\label{FGform}
\mathrm{d}s^2 = \frac{L^2}{r^2}\mathrm{d}r^2 +\frac{r^2}{L^2}\eta_{ab}E^a(r,\mathrm{x})E^b(r,\mathrm{x})\,.
\end{equation}
For torsionless connections there is always a suitable gauge choice 
such that the metrics (\ref{FGform}) are fully determined by two coefficients $\hat{e}^a$ and $\hat{f}^a$ in the 
expansion of the coframe one-forms $\hat{E}^a(r,\mathrm{x})$ along the holographic coordinate $r\in \mathbb{R}_+$
\begin{equation}
\label{vielbeil}
\hat {E}^a(r,\mathrm{x})= \left[\hat{e}^a(\mathrm{x})+\frac{ L^2}{r^2} \hat{F}^a(\mathrm{x})+\cdots\right]+\frac{ L^3}{r^3}\left[\hat{f}^a(\mathrm{x})+\cdots \right]\,.
\end{equation}
The asymptotic boundary is at   $r\to\infty$ and it is endowed with the geometry
\begin{equation}
\label{bry}
\mathrm{d}s^2_{\mathrm{bry.}}=\lim_{r\to \infty}\frac{\mathrm{d}s^2}{k^2r^2}.
\end{equation}
The ellipses in (\ref{vielbeil}) denote terms that are multiplied by higher negative powers of $r$. 
Their coefficients 
are determined by $\hat{e}^a$ and $\hat{f}^a$, and have specific geometrical interpretations\footnote{For example, the coefficient $\hat{F}^a$ is related to the boundary Schouten tensor.}, though this is not relevant for our discussion.

The $3+1$-split formalism makes clear that $\hat{e}^a(\mathrm{x})$ and $\hat{f}^a(\mathrm{x})$, being themselves vector-valued one-forms in the boundary, are the proper  canonical variables playing the role of boundary ``coordinate'' and ``momentum'' for the (hyperbolic) Hamiltonian evolution along $r$. For the stationary backgrounds under consideration, describing thermally equilibrated non-dissipating boundary fluid configurations, $\hat{e}^a$ and $\hat{f}^a$ are $t$-independent.   

The boundary ``coordinate'' is given by the set of one-forms $\hat{e}^a$. 
 For this coframe we must  determine the ``momentum'' of the boundary data. For example, when the boundary data carry zero mass, we expect this to be zero. In this case $\hat{f}^a(x)=0$ and the unique exact solution of the Einstein's equations is pure AdS$_4$. 

More generally, the vector-valued one-form $\hat{f}^a$ satisfies 
\begin{equation}
\label{condfa}
\hat{f}^a\wedge \hat{e}_a=0\,,\quad\epsilon_{abc}\hat{f}^a\wedge \hat{e}^b\wedge \hat{e}^c=0\,,\quad \epsilon_{abc}\mathrm{D}\hat{f}^b\wedge \hat{e}^c=0\,,
\end{equation}
where the action of the generalized exterior derivative $\mathrm{D}$ on a vector-valued one-form $\hat{V}^a$ is defined as
\begin{equation}
\label{covD}
\mathrm{D}\hat{V}^a=\mathrm{d}\hat{V}^a +\epsilon^{a}_{\hphantom{a}bc}\hat{B}^b\wedge\hat{e}^c\,,
\end{equation}
and the ``magnetic field'' $\hat{B}^a$ is the Levi--Civita spin connection associated with $\hat{e}^a$ \cite{Mansi:2008br}.
One can easily see that  conditions (\ref{condfa}) imply, respectively,  symmetry, absence of trace and covariant conservation of the 
tensor $T=T^a_{\hphantom{a}b}\check{e}_a\otimes \hat{e}^b$, defined as 
\begin{equation}
\label{fT}
\hat{f}^a=\frac{1}{\kappa}T(\hat{e}^a)=\frac{1}{\kappa}T^a_{\hphantom{a}b}\hat{e}^b
\,,\quad \kappa=\frac{3}{8\pi G_\mathrm{N} L}\, .
\end{equation}
Hence we can interpret the latter as the covariantly conserved energy--momentum tensor of a conformal field theory. Here we are interested in 
particular stationary bulk solutions for which we expect the energy--momentum tensor be reduced to the perfect relativistic form, Eqs. \eqref{e-m} and \eqref{perstr}: 
\begin{equation}
\label{fluidemST}
T^a_{\hphantom{a}b} = (\varepsilon +p) u^au_b +p\delta^a_{b}\,.
\end{equation}

Although the two necessary ingredients for the description of a relativistic perfect fluid, namely the boundary frame and the velocity one-form, are nicely packaged in 
the leading and subleading independent boundary data, 
until now we did not assume any specific relationship between them.  Nevertheless it is clear that such a relationship would be imposed by any {\it exact} solution 
of the bulk gravitational equations, given the interior boundary conditions. We will soon observe that the
 Fefferman--Graham expansion of the exact solutions of Sec. \ref{geoflu}
 yield the {\it same form} for the boundary 
energy--momentum tensor, namely
\begin{equation}
\label{efrel}
\hat{f}^0=-\frac{2M}{3L}\hat{e}^0\,,\quad \hat{f}^\alpha=\frac{M}{3L}\hat{e}^\alpha\,.
\end{equation}
The boundary frame one-forms $\hat{e}^a$ are themselves, of course, different in the three solutions. Comparing (\ref{fT}), (\ref{fluidemST}) and (\ref{efrel}), we find 
\begin{equation}\label{conpress}
\varepsilon = 2p=2\kappa\frac{M}{3L}\,,
\end{equation}
constant as already advertised.
The solutions under consideration describe thus the  \emph{same conformal fluid} in different kinematic states. More importantly, (\ref{efrel}) fixes the direction of the velocity field with respect to the boundary frame to be 
\begin{equation}
\label{comframe}
\check{u}=\check{e}_0\,.
\end{equation}
As explained in Sec. \ref{Papasec}, in the Papapetrou--Randers geometry (Eq.~(\ref{Papa})), this congruence is tangent to a constant-norm Killing field
and has thus zero shear, expansion and acceleration  (consistent, according to \eqref{perfeom-dissless}, with the constant pressure found in \eqref{conpress}). 
It also shows that the observer's frame $\check{e}_a$ is {\it comoving}.  Therefore, in the Fefferman--Graham expansion the kinematic properties of holographic 
fluids are determined by the geometric properties of the boundary  comoving frame. 



\subsection{Some exact geometries and their fluid interpretation}\label{geoflu}


We present here three examples of holographic fluids with vorticity that reside at the boundary of two exact solutions of the bulk vacuum Einstein equations; 
the Kerr--AdS$_4$, the Taub--NUT--AdS$_4$  and the hyperbolic NUT--AdS$_4$ black hole solutions. 

The four-dimensional Kerr solution of Einstein's equation with 
$\Lambda=-3k^2$
reads: 
\begin{eqnarray}
 \mathrm{d}s^2 = \frac{\mathrm{d}r^2}{V(r, \theta)}  -V(r, \theta)
      \left[ \mathrm{d}t - \frac{a}{\Xi} \sin^2\theta\, \mathrm{d}\phi\right]^2
  + \frac{\rho^2}{\Delta_\theta} \mathrm{d}\theta^2
  + \frac{\sin^2\theta \Delta_\theta}{\rho^2}\left[a\, \mathrm{d}t - 
      \frac{r^2+a^2}{\Xi}\,  \mathrm{d}\phi \right]^2,
      \label{KAdS}
\end{eqnarray}
where
\begin{equation}
V(r,\theta)=\frac{ \Delta_r }{\rho^2}
\end{equation}
and
\begin{equation}\label{KXi}
\begin{cases}
 \Delta_r  =  (r^2 + a^2)(1+ k^2 r^2) - 2Mr \\
  \rho^2  =  r^2 + a^2\cos^2\theta
\end{cases}
\qquad
\begin{cases}
 \Delta_\theta  =  1 - k^2 a^2 \cos^2\theta  \\
 \Xi  =  1 - k^2 a^2.
\end{cases}
\end{equation}
The geometry has inner ($r_-$) and outer ($r_+$) horizons, where  $ \Delta_r$ vanishes,  as well as an ergosphere at $g_{tt}=0$. 
The solution at hand describes the field generated by a mass $M$ rotating with an angular velocity 
\begin{equation}
\label{Kerr-ang-vel}
\Omega= \frac{a(1+ k^2 r_+^2)}{r^2_++a^2}
\end{equation}
as measured by a static observer at infinity \cite{Caldarelli:1999xj,Gibbons:2004ai}. Note that 
asymptotic observer associated with a natural frame of the coordinate system at hand is not static, but has an angular velocity
\begin{equation}
\Omega_{\infty} = a k^2.
\end{equation}
The boundary metric of Eq. \eqref{KAdS} is \eqref{KADS-bound} (without primes in the angular coordinates).

The Taub--NUT--AdS$_4$ geometry is a foliation over squashed three-spheres solving Einstein's equations with negative cosmological constant (the $\sigma$s are given in \eqref{LMC}):
\begin{eqnarray}
\mathrm{d}s^2 &=&
 \frac{\mathrm{d}r^2}{V(r)}
+\left(r^2+n^2\right)\left((\sigma^1)^2+
\left(\sigma^2\right)^2\right)-4n^2V(r)\left(\sigma^3\right)^2
\label{bulmet}
\\
\label{hyptaubnutpot-lor-t}
&=&   \frac{\mathrm{d}r^2}{V(r)}
+\left(r^2+n^2\right)\left(
 \mathrm{d}\theta^2 +
\sin^2\theta\, \mathrm{d}\phi^2
\right)-V(r)\left[\mathrm{d}t + 4n\sin^2 \frac{\theta}{2}\,  \mathrm{d}\phi\right]^2,
\end{eqnarray}
where
\begin{equation}
\label{potTN}
V(r)= \frac{\Delta_r}{\rho^2}
\end{equation}
and
\begin{equation}
\label{hyptaubnutpot-lor}
\begin{cases}
\Delta_r=r^2-n^2 -2 M r +k^2\left(r^4+6n^2r^2 -3 n^4\right)\\
\rho^2= r^2+n^2.
\end{cases}
\end{equation}
Besides the mass $M$ and the cosmological constant $\Lambda=-3k^2$, this solution depends on an extra parameter $n$: the nut charge. 

The solution at hand has generically two horizons ($V(r_\pm)=0$) and is well-defined outside the outer horizon $r_+$, where $V(r)>0$. 
The nut is the endpoint of a Misner string \cite{misner:1963}, departing from $r=r_+$, all the way to $r\to\infty$, on the southern pole at $\theta=\pi$. The geometry 
is nowhere singular along the Misner string, which appears as a coordinate artifact much like the Dirac string of  a magnetic monopole is a gauge artifact. 
In order for this string to be invisible, coordinate transformations displacing the string must be univalued everywhere, which is achieved by requiring the 
periodicity condition  $t\equiv t - 8\pi n$. 

One can avoid periodic time and keep the Misner string as part of the geometry. This semi-infinite spike appears then as a source of angular momentum, integrating to 
zero \cite{bonnor:1969, dowker:1974}, and movable at wish using the transformations generated by the above vectors. This will be our viewpoint throughout this work.
However, despite the non-compact time, the Taub--NUT--AdS geometry is plagued with closed timelike curves, which disappear only in the vacuum limit $k\to 0$ \cite{Astefanesei:2004kn}. Even though this is usually an unwanted situation, it is not sufficient for rejecting the geometry, which from the holographic perspective 
has many interesting and novel features. The boundary geometry of \eqref{hyptaubnutpot-lor-t}, reached by following \eqref{bry}, is \eqref{berg-lor-t} and this is the 
background where the fluid is evolving. What we called Misner point for the latter in Sec.~\ref{S3} is the endpoint of the bulk Misner string.  

On a non-compact horizon, the nut charge can be pushed to infinity. This happens in hyperbolic NUT black holes, obtained as foliations over three-dimensional anti-de Sitter spaces\footnote{In their Euclidean section these geometries have no nut, but only a bolt \cite{Chamblin:1998pz}. We shall nevertheless conform to standard use and call them -- with a slight abuse of language -- {\em hyperbolic NUT black holes}, to stress the presence of a non-trivial $S^1$ fibration over $H_2$. Physically, they represent {\em rotating black hyperboloid membranes} \cite{Caldarelli:2008pz}.}. Using the $SL(2,\mathrm{R})$ Maurer--Cartan forms \eqref{LMCSL}, we obtain the following solution of Einstein's equations with cosmological constant 
$\Lambda=-3k^2$:
\begin{eqnarray}
\mathrm{d}s^2 &=&
 \frac{\mathrm{d}r^2}{V(r)}
+\left(r^2+p^2\right)\left((\rho^1)^2+
\left(\rho^2\right)^2\right)-4p^2V(r)\left(\rho^0\right)^2
\label{bulmethyp}
\\
\label{hypnutpot-lor-t}
&=&   \frac{\mathrm{d}r^2}{V(r)}
+\left(r^2+p^2\right)\left(
 \mathrm{d}\sigma^2 +
\sinh^2\sigma\,\mathrm{d}\phi^2
\right)-V(r)\left[\mathrm{d}t-4p\sinh^2\frac{\sigma}{2}\,\mathrm{d}\phi\right]^2
\end{eqnarray}
with $V(r)$ given in \eqref{potTN} and 
\begin{equation}
\label{hypnutpot-lor}
\begin{cases}
\Delta_r=-r^2+p^2  -2\hat M r+k^2\left(r^4+6p^2r^2 -3 p^4\right)\\
\rho^2= r^2+p^2.
\end{cases}
\end{equation}
Here $\hat M$ is the mass parameter and $p$ characterizes the non-trivial $S^1$ fibration over the $H_2$ base. In this case no Misner string is however present, and the space is globally hyperbolic provided $p\leq\nicefrac{L}{2}$.
The boundary metric of \eqref{hypnutpot-lor-t} is \eqref{SLell}, which plays the role of host for the holographic fluid.
Interestingly, this family of solutions is connected to the Kerr--AdS$_4$ black hole, as we will now show.

The Kerr--AdS$_4$ black hole \eqref{KAdS} has a rotation parameter $a$ restricted to $a^2<L^2$, and is singular for $a^2=L^2$. It has however a finite, maximally spinning limit if the $a\rightarrow L$ limit is taken keeping the horizon size finite and simultaneously zooming into the pole \cite{Caldarelli:2008pz}. More explicitly, we trade the angle $\theta$ in \eqref{KAdS} for a new coordinate $\sigma$ according to
\begin{equation}
\sin\theta=\sqrt\Xi\,\sinh^2\nicefrac\sigma2.
\label{theta2sigma}\end{equation}
Then, the resulting metric has a regular $a\rightarrow L$ limit,
\begin{equation}
\mathrm{d}s^2=\frac{\mathrm{d}r^2}{V(r)}
+\frac{1+k^2r^2}{4k^2}\left(\mathrm{d}\sigma^2 +
\sinh^2\sigma \mathrm{d}\phi^2\right)
-V(r)\left[\mathrm{d}t - \frac1k\sinh^2 \frac{\sigma}{2}\,  \mathrm{d}\phi\right]^2,
\label{usKAdS}\end{equation}
with
\begin{equation}
V(r)=1+k^2r^2-\frac{2k^2Mr}{1+k^2r^2}.
\end{equation}
This is again a rotating black hyperboloid membrane. Indeed, performing the rescaling $r\mapsto2r$, $t\mapsto \nicefrac{t}{2}$, this metric is cast in the form \eqref{hypnutpot-lor-t}--\eqref{hypnutpot-lor} with parameters $\hat M=\nicefrac{M}{8}$ and $p=\nicefrac L2$, and the boundary of \eqref{usKAdS} takes the form \eqref{SLell} with this value of $p$. The boundary is therefore the undeformed AdS$_3$ spacetime that can equivalently be recovered by performing the coordinate transformation \eqref{theta2sigma} directly on the Kerr--AdS boundary metric \eqref{KADS-bound}, followed by the $a\rightarrow L\,$ limit.
As we saw, this special value corresponds to the limiting case that enjoys global hyperbolicity. Accordingly, from the Kerr--AdS point of view, it is obtained as the ultraspinning limit for which the boundary Einstein static universe rotates effectively at the speed of light \cite{Hawking:1998kw}. In this way, we nicely connected fluids on warped AdS$_3$ backgrounds to fluids living on a rotating Einstein static universe into a single, continuous family.

There are many other exact bulk four-dimensional geometries that one could study under the perspective of describing boundary fluids with vorticity. 
One can for instance consider the case of flat horizon reached when the trigonometric sinus in \eqref{hyptaubnutpot-lor-t} 
(or the hyperbolic sinus in \eqref{hypnutpot-lor-t}) is traded for a linear function and the potential adapted by 
dropping the first two terms (see e.g.~\cite{Astefanesei:2004kn, Chamblin:1998pz}). This bulk solution, also
plagued by the global hyperbolicity problem, gives rise to a boundary fluid moving on the Som--Raychaudhuri 
geometry \eqref{KAdS-bry-pol}. Alternatively, the flat-horizon bulk solution can be obtained as an appropriate 
pole-zooming of the four-dimensional Kerr--AdS, consistent with the observed relationships among the boundary 
geometries. Limits at $n,p\to 0$ lead to the so-called topological black holes \cite{Mann:1996gj,Vanzo:1997gw,Brill:1997mf}. These are interesting in their 
own right \cite{Emp99} even though the holographic fluid dynamics has no intrinsic vorticity -- their 
boundaries are Einstein static universes $\mathbb{R}\times S^2$ given in \eqref{ESU} or $\mathbb{R}\times H_2$,  
where no fiber appears that would create vorticity. Hyperbolic Kerr--AdS or other exact bulk metrics can be 
found to reproduce on the boundary \eqref{ERUhyp-bound} or \eqref{AKT-south} \cite{Klemm:1997ea}. One can also find solutions 
that combine nut charge and ordinary rotation \cite{demianski:1966} such as Kerr--Taub--NUT--AdS\@. We will neither pursue 
any longer the general analysis of this rich web of backgrounds exhibiting many interrelations, nor delve into a 
quantitative presentation, but move instead into another interesting approach to holographic fluids, which can be easily exemplified with the backgrounds at hand. 


The above bulk geometries describe holographically a conformal fluid at rest without shear and expansion in the 
Papapetrou--Randers frame of a Papapetrou--Randers geometry \eqref{Papa}. These fluids have a non-trivial kinematics, 
though, because of the vorticity of the geodesic congruence they fill. The vorticity is different in the various cases:
it is given in Eqs. \eqref{Kerrvort}, \eqref{TNvort} and \eqref{hypvort}, for Kerr--AdS, Taub--NUT--AdS and hyperbolic 
NUT--AdS\@. In the first case, the fluid undergoes a cyclonic motion with maximal vorticity at the poles and vanishing at the equator. 
In the other two backgrounds, the vorticity is constant as a consequence of the homogeneity. The velocity fields are not homogeneous,
though, and behave differently in Taub--NUT--AdS and hyperbolic NUT--AdS\@.

Even though the boundary spacetime of Taub--NUT--AdS is homogeneous, the constant-$t$ surfaces are not.  
Inertial observers, comoving with the fluid have therefore a different perception  depending on whether they 
are at $0<\theta<\theta_*$ or in the disk $\theta_*<\theta<\pi$, surrounding the Misner string. This gives a 
physical existence to the $b^2=1$ edge,
the meaning of which is better expressed in the Zermelo frame. In the latter, 
the fluid becomes superluminal and the Misner string is interpreted  as the core of the vortex with homogeneous vorticity.

The various troublesome features which appear in G\"odel-like spaces as the ones at hand 
are intimately related with the non-trivial rotational properties combined with the homogeneous character of these manifolds. In other words, for the Taub--NUT--AdS boundary, 
they are due to the existence of a monopole-like Misner vortex\footnote{Since the bulk theory is such that the boundary does not have access to a charge 
current, the Misner vortex cannot be associated with a vortex in an ordinary superfluid, but is related to the spinning string of \cite{Mazur:1986gb}, 
the metric of which, Eq.~(\ref{AKT-south}), indeed appears when zooming in on the southern pole.}. 
Although no satisfactory physical meaning has ever  been given to G\"odel-like spaces, the causal consistency of the latter being still questionable, 
they seem from our holographic perspective to admit a sensible interpretation in terms of conformal fluids evolving in homogeneous vortices \eqref{TNvort}\footnote{As already stressed,  one should  add a $\delta$-function contribution to the Taub--NUT--AdS vorticity \eqref{TNvort} because we keep the Misner string physical with non-compact time \cite{Leigh:2011au,LPP2}.} or \eqref{hypvort}.

The case of hyperbolic NUT--AdS, Eq. \eqref{hypnutpot-lor-t}, is yet of a different nature. This bulk geometry leads again to homogeneous 
boundary \eqref{SLell}. Hence, the fluid has constant vorticity \eqref{hypvort}. However, in the case at hand, the spatial sections 
$\mathrm{d}x^2=a_{ij}\mathrm{d}x^i\mathrm{d}x^j$ are non-compact and negatively curved as opposed to the boundary of Taub--NUT--AdS\@. 
As a consequence, the combination of vorticity and homogeneity does not break global hyperbolicity, as long as $p\leq \nicefrac{L}{2}$. In 
this regime, the velocity of the fluid is well-defined everywhere, and its Lorentz factor with respect to Zermelo observer  is increasing with 
$\sigma$ and bounded as $\gamma\leq\nicefrac{1}{\sqrt{1-4p^2k^2}}$. 
For this observer,
the fluid is at rest in the center (\emph{i.e.}~at the north pole) and fast rotating 
at infinity. 
When $p= \nicefrac{L}{2}$, it reaches the speed of light when $\sigma \to \infty$, whereas for $p> \nicefrac{L}{2}$ 
this happens at finite $\sigma=\sigma_*$, along the surface-of-light edge. The latter situation is similar to what happens in the Taub--NUT--AdS irrespectively of the value of the nut charge $n$. In the hyperbolic case, the major differerence is however that the vortex, together with the Misner point are sent to spatial infinity ($\sigma\to\infty$).

The above discussion holds in the perspective of interpreting the holographic data as a genuine stationary fluid. There is however an alternative viewpoint already advertised, consisting in the analogue gravity interpretation of the boundary gravitational  background. From the latter, the physical data are still ($h_{ij}, W^i$) \emph{i.e.}~a two-dimensional geometry and a velocity field.
However, their combination into (\ref{Zermelo}) is not a physical spacetime. The would be light cone, in particular, is narrowed down to the sound or light velocities in the medium under consideration -- necessarily smaller than the velocity of light in vacuum. Consequently, the breaking of hyperbolicity or the appearance of CTCs are not issues of concern, and the 
regions where $\gamma$ becomes imaginary keep having a satisfactory physical interpretation as portions of space, where the medium is supersonic/superluminal  with respect to the sound/light velocity \emph{in the medium and not in the vacuum}. Finally, the virtual spacetime (\ref{Zermelo}) governs the mode propagation through the fluid. This way of thinking opens up a new chapter that requires adjusting suitably the standard holographic dictionary. The latter provides indirect information on the physical system that must be retrieved. 




\section{Alternative expansion: from $\mathbf{2+1}$ to $\mathbf{3+1}$}\label{Min}

The holographic fluids we have described so far emerge from exact bulk four-dimensional solutions of vacuum Einstein's equations with negative cosmological constant. They appear as a set of two pieces of boundary data -- the coframe and the energy--momentum current -- following the Fefferman--Graham expansion at large (appropriately chosen) radial coordinate. This is a top-down approach as opposed to the alternative bottom-up method initiated in \cite{Hubeny}.  The latter aims at reconstructing perturbatively a bulk solution starting from boundary data. 
The perturbative expansion of the bulk geometry obtained in this way is, however, orthogonal in spirit to that of the Fefferman--Graham series, since it is a derivative (of the velocity field) rather than a large-radius expansion. It captures therefore from the very first order the presence of 
a regular horizon of the black object that generates the dynamics of the boundary fluid.

As a matter of principle, one could generally follow the above procedure and perturbatively reconstruct the bulk solution corresponding to any boundary background of the Papapetrou--Randers type in terms of the data $(b_i, a_{ij})$ containing the full dynamics of the fluid. Our viewpoint is different though, and we are here interested in modestly discussing the interplay between the perturbative expansion 
developed in \cite{Hubeny,Bhattacharyya:2008ji, Bhattacharyya:20082} and the exact solutions we have analyzed in Sec.~\ref{holog}. This is motivated by the observation made in \cite{Bhattacharyya:2008ji, Bhattacharyya:20082}, according to which the proposed bulk perturbative reconstruction of the Kerr--AdS boundary fluid (in several dimensions) does coincide exactly with the original bulk geometry at second order -- modulo a specific resummation, indicative of the genuinely all-order nature of the solution. This is remarkable and leads to the deeper question: why and under which conditions does this occur?

The question raised here is twofold. Given an exact bulk solution, what can make it be expressed in the form of a limited expansion in terms of its boundary data  obtained via the Fefferman--Graham procedure? Given a set of arbitrary boundary data, what could ensure the corresponding bulk series be exact at finite order? 

Making progress in this direction would require delving into the physics of dissipative phenomena and their holographic expressions. This analysis stands beyond our present scope. We can nevertheless make an observation that might ultimately be relevant. In all backgrounds under consideration, the bulk can be expressed exactly as a limited derivative expansion provided an extra term  (with respect to the expansion proposed in \cite{Bhattacharyya:2008ji, Bhattacharyya:20082}) involving the Cotton tensor of the boundary geometry is appropriately added, and after performing a resummation similar to that of Kerr--AdS\@. 
This holds for all Papapetrou--Randers  backgrounds presented in Sec.~\ref{ex21}. To keep our presentation compact, we will only consider those for which we studied the bulk realisation in Sec.~\ref{holog}, namely Kerr--AdS, Taub--NUT--AdS and hyperbolic NUT--AdS\@. 

The starting point for this analysis is the expression in Eddington--Finkelstein coordinates of the bulk metrics. For Kerr--AdS, Eq.~\eqref{KAdS} this is achieved by performing the following coordinate change:
\begin{equation}
\label{kerrEF}
\begin{cases}
\mathrm{d}t\mapsto \mathrm{d}t -\frac{r^2 + a^2}{\Delta_r}\mathrm{d}r
 \\
\mathrm{d}\phi\mapsto\mathrm{d}\phi -\frac{a\Xi}{\Delta_r}\mathrm{d}r
\end{cases}
\end{equation}
with all quantities defined in \eqref{hyptaubnutpot-lor}. Similarly for Taub--NUT--AdS, Eqs.  \eqref{hyptaubnutpot-lor-t} and \eqref{hyptaubnutpot-lor}, one performs 
\begin{equation}
\label{TN-hypEF}
\mathrm{d}t\mapsto \mathrm{d}t -\frac{r^2 + n^2}{\Delta_r}\mathrm{d}r,
\end{equation}
while the same holds with $n\mapsto p$ for the hyperbolic NUT--AdS given in \eqref{hypnutpot-lor-t} and \eqref{hypnutpot-lor}. All three bulk metrics assume then the following generic form:
\begin{equation}
\label{papaef}
\mathrm{d}s^2 =
-2\hat{u}\mathrm{d}r+r^2k^2\mathrm{d}s^2_{\mathrm{bry.}}+\frac{1}{k^2}\Sigma_{\mu\nu} 
\mathrm{d}x^\mu\mathrm{d}x^\nu
+ \frac{\hat{u}^2}{\rho^2}  \left(2Mr+\frac{u^\lambda C_{\lambda \mu}\epsilon^{\mu\nu\sigma}\omega_{\nu\sigma}}{2k^6\sqrt{-g_{\mathrm{bry.}}}}\right),
\end{equation}
where all the quantities refer to the boundary metric $\mathrm{d}s^2_{\mathrm{bry.}}$. The latter being of the Papapetrou--Randers form \eqref{Papa}, 
$\hat{u}=-\mathrm{d}t+b$ and $\omega=\frac{1}{2}\mathrm{d}b$. Furthermore  
$C_{\lambda \mu}$ are the components of the Cotton tensor (zero for Kerr--AdS, and displayed in Eqs. \eqref{CotSU} and \eqref{CotSL} for the other cases). Finally
\begin{eqnarray}
\Sigma_{\mu\nu} 
\mathrm{d}x^\mu\mathrm{d}x^\nu&=&-2\hat{u}\nabla_\nu \omega^\nu_{\hphantom{\nu}\mu}\mathrm{d}x^\mu- \omega_\mu^{\hphantom{\mu}\lambda} \omega^{\vphantom{\lambda}}_{\lambda\nu}\mathrm{d}x^\mu\mathrm{d}x^\nu
-\hat{u}^2\frac{R}{2},\\ \label{rho2}
\rho^2&=& r^2 +\frac{1}{2k^4} \omega_{\mu\nu} \omega^{\mu\nu},
\end{eqnarray}
and $\rho^2$, as computed in \eqref{rho2}, coincides with the quantities defined in Eqs.
\eqref{KXi}, \eqref{hyptaubnutpot-lor} and \eqref{hypnutpot-lor}.

The above result \eqref{papaef} deserves a discussion. It appears as a limited derivative expansion on the velocity field of the boundary geodesic congruence $\check{u}=\partial_t$. The latter carries neither expansion, nor shear (see Secs.~\ref{vfc} and \ref{Papasec}) but only vorticity given in Eqs. \eqref{Kerrvort},  \eqref{TNvort} and  \eqref{hypvort} for the three backgrounds under investigation Kerr--AdS \eqref{KAdS}, Taub--NUT--AdS \eqref{hyptaubnutpot-lor-t} and the hyperbolic NUT--AdS \eqref{hypnutpot-lor-t}. It seems that at most two derivatives of the velocity field are involved, but this counting is naive. Indeed, the vorticity being ultimately an intrinsic property of the geometry\footnote{Vorticity components are directly related to the connection coefficients, as e.g.~$\Gamma^i_{tj}=-\omega^i_{\hphantom{i}j}$.}, $R\sim \omega^2$, while $C\omega\sim \omega^4$. Furthermore, $\nicefrac{1}{\rho^2}$ is a resummed power series in even powers of the vorticity. As already advertised, this resummation betrays the infinite perturbative expansion underlying the method, that would otherwise appear as limited to the fourth order.

The metric \eqref{papaef} yields an exact solution of AdS$_4$ gravity for a large class of boundary Randers data $(b_i,a_{ij})$. In addition to the cases described above, one can rewrite in this form the full Kerr--Taub--NUT--AdS$_4$ family of metrics, as well as all rotating topological black holes found in \cite{Klemm:1997ea}: the rotating black cylinder and the rotating hyperbolic black membrane. All these metrics belong to the Pleba\'nski--Demia\'nski type-D class of solutions \cite{Plebanski:1976gy}. It is an interesting problem to see if it is possible to extend this collection and find the most general set of Randers data $(b_i,a_{ij})$ generating an exact solution through \eqref{papaef}.

It is remarkable that all known exact AdS$_4$ black hole solutions can be set in the above form \eqref{papaef}, much like Kerr--AdS,
provided an extra term (one should say an extra resummed series) based on the Cotton tensor is added. This term, of fourth order in the derivatives of the velocity field\footnote{The Cotton tensor itself is third order in the derivatives of the velocity field, but it cannot appear at this order in the fluid/gravity metric because it has the wrong parity. The fourth order is indeed the smallest order for which it can appear.}, was absent in the original expressions of \cite{Bhattacharyya:2008ji, Bhattacharyya:20082}, only valid up to second order. In five or higher bulk dimensions, terms involving the Weyl tensor appear at the second order \cite{Bhattacharyya:2008ji,Bhattacharyya:20082}, but obviously do not contribute in the four-dimensional case under consideration. Our expression \eqref{papaef} shows that in four dimensions analogous terms, involving the Cotton tensor, appear in the derivative expansion starting from the fourth order.

\section{Conclusions}\label{con}

In this review we presented an extensive discussion of the holographic description of vorticity. This is the first step in efforts to extend AdS/CMT to systems such a rotating atomic gases of analogue gravity systems. The upshot is that even the simplest setup, namely non-dissipating fluids in local equilibrium with non-zero vorticity, has an extremely rich geometric structure whose detailed analysis should lead to new and interesting physical results. One such result, presented in \cite{LPP2} was the calculation of the classical rotational Hall viscosity coefficient of neutral $2+1$ dimensional fluids having uniform vorticity $\Omega$ 
\begin{equation}
\zeta_\mathrm{H}=\frac{\varepsilon +p}{\Omega},
\end{equation}
which we were not able to find in recent works on parity broken hydrodynamics in $2+1$ dimensions.

The next steps in our program will certainly reveal interesting physical consequences. For example, the study of scalar, vector and ultimately graviton fluctuations around the above geometries should lead to the determination of various transport coefficients for rotating neutral fluids. These developments might also shed light on the thermalization processes that are expected near analogue horizons. Furthermore, we believe that our approach offers a well-defined path to study the issue of time-dependence in conjunction with irreversible, non-equilibrium dynamics
as it can appear in dissipative fluid configurations or in the vicinity of analogue horizons.

In the present work we have emphasized the importance of the nut charge in the holographic description of vorticity. In superfluids this is a quantized quantity, hence one might wonder whether and how nut-charge quantization could emerge in their holographic description.  The latter is in fact incomplete and the formation of vortices in rotating condensates calls for a more complete understanding, which justifies our efforts.

An  issue worth mentioning is the relationship of our work with alternative approaches of fluid/gravity correspondence. We have presented some preliminary results in Sec. \ref{Min} and we plan to elaborate on that subject in a forthcoming work. Many other roads  seem open for further investigation, which we have not discussed. One could for example try to describe fluids in more complicated  kinematic states, with multipolar vorticity -- as a generalization of the monopole-like configurations created by nut charges, or the dipoles corresponding to Kerr cyclonic motions. This would require the generalization of the Weyl multipole solutions to asymptotically AdS spaces. The magnetic paradigm of the geodesic motion in Papapetrou--Randers geometries (see e.g. \cite{Gibbons}) might turn in a powerful tool for that task. Let us finally mention that analogue-gravity applications are very rich and diverse. Setting the bridge with holographic techniques  would however require a more systematic study.

\section*{Acknowledgements}
The authors benefited from discussions with  C.~Bachas, C.~Charmousis, S.~Katmadas, R.~Mey- er, V.~Niarchos, G.~Policastro and K.~Sfetsos. P.M.P.~and K.S.~would like to thank the University of Crete, the University of Ioannina and the University of Patras, and R.G.L.~and A.C.P.~thank the CPHT of Ecole Polytechnique for hospitality. 
This research was supported by the LABEX P2IO, the ANR contract  05-BLAN-NT09-573739, the ERC Advanced Grant  226371, the ITN programme PITN-GA-2009-237920  , the IFCPAR CEFIPRA programme 4104-2, the U.S. Department of Energy contract FG02-91-ER4070 and the EU contract FP7-REGPOT-2008-1: CreteHEPCosmo-228644. It was also cofinanced by the European Union (European Social Fund, ESF) and Greek national funds through the Operational Program Education and Lifelong Learning" of the National Strategic Reference Framework (NSRF) under Funding of proposals that have received a positive evaluation in the 3rd and 4th Call of ERC Grant Schemes.


\begin{thebibliography}{99}

\bibitem{Damour:1979}
T. Damour, ``Quelques propri\'et\'es m\'ecaniques, \'electromagn\'etiques, thermodynamiques et quantiques des trous noirs,'' Th\`ese de Doctorat d'Etat, Universit\'e Pierre et Marie Curie, Paris VI, 1979.

\bibitem{Hubeny}
  V.E.~Hubeny, S.~Minwalla and M.~Rangamani,
  ``The fluid/gravity correspondence,''
  [arXiv:1107.5780 [hep-th]].

\bibitem{Hartnoll:2009sz}
  S.A.~Hartnoll,
  ``Lectures on holographic methods for condensed matter physics,''
  Class.\ Quant.\ Grav.\  {\bf 26} (2009) 224002
  [arXiv:0903.3246 [hep-th]].

\bibitem{Herzog:2009xv}
  C.P.~Herzog,
  ``Lectures on holographic superfluidity and superconductivity,''
  J.\ Phys.\  {\bf A42} (2009) 343001
  [arXiv:0904.1975 [hep-th]].

\bibitem{Horowitz:2010gk}
  G.T.~Horowitz,
  ``Introduction to holographic superconductors,''
  arXiv:1002.1722 [hep-th].

\bibitem{Cooper}
N.R. Cooper, ``Rapidly rotating atomic gases,''  Adv. in Phys. \textbf{57} (2008) 539.

\bibitem{Fetter}
A.L. Fetter, ``Rotating trapped Bose--Einstein condensates,'' 
Rev. Mod. Phys. \textbf{81} (2009) 647.

\bibitem{Leonhardt} 
U. Leonhardt and P. Piwnicki, ``Relativistic effects of light in moving media with extremely low group velocity,'' Phys.  Rev. Lett.  \textbf{84} (2000) 822.

\bibitem{Unruh1}
  W.G.~Unruh,
  ``Experimental black hole evaporation,''
  Phys.\ Rev.\ Lett.\  {\bf 46} (1981) 1351.

\bibitem{Unruh2}
  W.G.~Unruh,
  ``Sonic analog of black holes and the effects of high frequencies on black
  hole evaporation,''
  Phys.\ Rev.\  {\bf D51} (1995) 2827.
  

\bibitem{Barcelo}
  C.~Barcel\`o, S.~Liberati and M.~Visser,
 ``Analogue gravity,''
  Living Rev.\ Rel.\  {\bf 8} (2005) 12
  [gr-qc/0505065].

\bibitem{Liberati}
  S.~Liberati, A.~Prain and M.~Visser,
 ``Quantum vacuum radiation in optical glass,''
  [arXiv:1111.0214 [gr-qc]].

\bibitem{NewPaper}
M. Caldarelli, R.G. Leigh, A.C. Petkou, P.M. Petropoulos, V. Pozzoli and K. Siampos, to appear.

\bibitem{Ehlers:1993gf}
  J.~Ehlers,
  ``Contributions to the relativistic mechanics of continuous media,''
  Gen.\ Rel.\ Grav.\  {\bf 25} (1993) 1225.

\bibitem{vanElst:1996dr}
  H.~van Elst and C.~Uggla,
  ``General relativistic $1+3$-orthonormal frame approach revisited,''
  Class.\ Quant.\ Grav.\  {\bf 14} (1997) 2673
  [gr-qc/9603026].

\bibitem{Bhattacharyya:2008ji}
  S.~Bhattacharyya, R.~Loganayagam, S.~Minwalla, S.~Nampuri, S.P.~Trivedi and S.R.~Wadia,
  ``Forced fluid dynamics from gravity,''
  JHEP {\bf 0902} (2009) 018
  [arXiv:0806.0006 [hep-th]].

\bibitem{Caldarelli:2008mv}
  M.M.~Caldarelli, O.J.C.~Dias, R.~Emparan and D.~Klemm,
  ``Black holes as lumps of fluid,''
  JHEP {\bf 0904} (2009) 024
  [arXiv:0811.2381 [hep-th]].
  
\bibitem{Caldarelli:2008ze}
  M.M.~Caldarelli, O.J.C.~Dias and D.~Klemm,
  ``Dyonic AdS black holes from magnetohydrodynamics,''
  JHEP {\bf 0903} (2009) 025
  [arXiv:0812.0801 [hep-th]].
 
\bibitem{Papapetrou}
A. Papapetrou, ``Champs gravitationnels stationnaires \`a sym\'etrie axiale,"  Ann. Inst. H. Poincar\'e {\bf A4} (1966) 83.


\bibitem{Randers}
G. Randers, ``On an asymmetrical metric in the four-space of general relativity," Phys. Rev. {\bf 59}
(1941) 195.

\bibitem{Gibbons}
  G.W.~Gibbons, C.A.R.~Herdeiro, C.M.~Warnick and M.C.~Werner,
 ``Stationary metrics and optical Zermelo--Randers--Finsler geometry,''
  Phys.\ Rev.\ {\bf D79} (2009) 044022
  [arXiv:0811.2877 [gr-qc]].

\bibitem{Leigh:2011au}
  R.G.~Leigh, A.C.~Petkou and P.M.~Petropoulos,
  ``Holographic three-dimensional fluids with nontrivial vorticity,''
  Phys.\ Rev. {\bf D85} (2012) 086010
  [arXiv:1108.1393 [hep-th]].


\bibitem{LPP2}
  R.G.~Leigh, A.C.~Petkou and P.M.~Petropoulos,
  ``Holographic fluids with vorticity and analogue gravity systems,'' 
   [arXiv:1205.6140 [hep-th]].

\bibitem{Zer31}
E. Zermelo, ``\"Uber das Navigationsproblem bei ruhender oder ver\"anderlicher Windverteilung,'' Z. Angew. Math. Mech. \textbf{11} (1931) 114.

\bibitem{Bardeen} 
J.M. Bardeen, ``A variational principle for rotating stars in general relativity,'' Ast. Journ. \textbf{162} (1970) 71. 

\bibitem{Banados:1992wn}
  M.~Ba\~nados, C.~Teitelboim and J.~Zanelli,
  ``The Black hole in three-dimensional space--time,''
  Phys.\ Rev.\ Lett.\  {\bf 69} (1992) 1849
  [hep-th/9204099].
  
\bibitem{Unruh:1980cg}
  W.G.~Unruh,
  ``Experimental black hole evaporation,''
  Phys.\ Rev.\ Lett.\  {\bf 46} (1981) 1351.

\bibitem{Unruh:1994je}
  W.G.~Unruh,
  ``Sonic analog of black holes and the effects of high frequencies on black
  hole evaporation,''
  Phys.\ Rev.\  {\bf D51} (1995) 2827.

  
\bibitem{Cacciatori}

 S.L.~Cacciatori, F.~Belgiorno, V.~Gorini, G.~Ortenzi, L.~Rizzi, V.G.~Sala and D.~Faccio,
 ``Space--time geometries and light trapping in traveling refractive index perturbations,''
  New J.\ Phys.\  {\bf 12} (2010) 095021
  [arXiv:1006.1097 [physics.optics]].



\bibitem{Shapere}
  S.R.~Das, A.~Ghosh, J.-H.~Oh and A.D.~Shapere,
 ``On dumb holes and their gravity duals,''
  JHEP {\bf 1104} (2011) 030
  [arXiv:1011.3822 [hep-th]].

\bibitem{Milnor}
J. Milnor, ``Curvatures of left-invariant metrics on Lie groups'', Advances in Math. \textbf{21} (1976) 293.

\bibitem{Scott}
P. Scott, ``The geometries of 3-manifolds'', Bull. London Math. Soc. \textbf{15} (1983) 401.


   \bibitem{rayPRD80}
  A.K. Raychaudhuri and S.N. Guha Thakurta,
  ``Homogeneous space--times of the G\"odel type,'' Phys. Rev. \textbf{D22} (1980) 802.
  
      \bibitem{rebPRD83}
M.J. Rebou\c{c}as and J. Tiomno,
  ``Homogeneity of Riemannian space--times of G\"odel type,'' Phys. Rev. \textbf{D28} (1983) 1251.

\bibitem{Gibbons:1999uv}
  G.~W.~Gibbons and C.~A.~R.~Herdeiro,
  ``Supersymmetric rotating black holes and causality violation,''
  Class.\ Quant.\ Grav.\  {\bf 16} (1999) 3619
  [hep-th/9906098].
 
\bibitem{Drukker:2003mg}
  N.~Drukker, B.~Fiol and J.~Simon,
  ``G\"odel type universes and the Landau problem,''
  JCAP {\bf 0410} (2004) 012
  [hep-th/0309199].
 
\bibitem{Hikida:2003yd}
  Y.~Hikida and S.J.~Rey,
  ``Can branes travel beyond CTC horizon in G\"odel universe?,''
  Nucl.\ Phys.\ B {\bf 669} (2003) 57
  [hep-th/0306148].

\bibitem{Israel:2003cx}
  D.~Israel,
  ``Quantization of heterotic strings in a G\"odel / anti-de Sitter spacetime and chronology protection,''
  JHEP {\bf 0401} (2004) 042
  [hep-th/0310158].
 
\bibitem{Israel:2004vv}
  D.~Israel, C.~Kounnas, D.~Orlando and P.M.~Petropoulos,
  ``Electric/magnetic deformations of $S^3$ and AdS$_3$, and geometric cosets,''
  Fortsch.\ Phys.\  {\bf 53} (2005) 73
  [hep-th/0405213].

\bibitem{Israel:2004cd}
  D.~Israel, C.~Kounnas, D.~Orlando and P.M.~Petropoulos,
  ``Heterotic strings on homogeneous spaces,''
  Fortsch.\ Phys.\  {\bf 53} (2005) 1030
  [hep-th/0412220].

\bibitem{Caldarelli:2004mz}
  M.M.~Caldarelli, D.~Klemm and P.J.~Silva,
  ``Chronology protection in anti-de Sitter,''
  Class.\ Quant.\ Grav.\  {\bf 22} (2005) 3461
  [hep-th/0411203].

\bibitem{cs}
S. Deser, R. Jackiw and S. Templeton, ``Topologically massive gauge theories,''
Ann. Phys. \textbf{140} (1982) 372; Erratum-ibid. \textbf{185}
(1988) 406; ``Three-dimensional massive gauge theories,'' Phys. Rev. Lett.
\textbf{48} (1982) 975.

\bibitem{Banados:2005da}
  M.~Ba\~nados, G.~Barnich, G.~Comp\`ere and A.~Gomberoff,
  ``Three-dimensional origin of G\"odel spacetimes and black holes,''
  Phys.\ Rev.\ {\bf D73} (2006) 044006
  [hep-th/0512105].

\bibitem{Anninos:2008fx}
  D.~Anninos, W.~Li, M.~Padi, W.~Song and A.~Strominger,
  ``Warped AdS$_3$ black holes,''
  JHEP {\bf 0903} (2009) 130
  [arXiv:0807.3040 [hep-th]].

\bibitem{Anninos:2011vd}
  D.~Anninos, S.~de Buyl and S.~Detournay,
  ``Holography for a de Sitter--Esque geometry,''
  JHEP {\bf 1105} (2011) 003
  [arXiv:1102.3178 [hep-th]].

\bibitem{KS65}
R.P. Kerr and A Schild, ``Some algebraically degenerate solutions of Einstein's gravitational field equations'', Proc. Symp. Appl. Math. \textbf{17} (1965) 199.



\bibitem{Rooman:1998xf}
  M.~Rooman and P.~Spindel,
  ``G\"odel metric as a squashed anti-de Sitter geometry,''
  Class.\ Quant.\ Grav.\  {\bf 15} (1998) 3241
  [gr-qc/9804027].


\bibitem{Hawking:1998kw}
  S.W.~Hawking, C.J.~Hunter and M.~Taylor,
  ``Rotation and the AdS/CFT correspondence,''
  Phys.\ Rev.\   {\bf D59} (1999) 064005
  [arXiv:hep-th/9811056].

     \bibitem{SR68}
  M.M. Som and  A.K. Raychaudhuri, ``Cylindrically symmetric charged dust distribution in rigid rotation in general relativity,'' Proc. R. Soc. London \textbf{A304} (1968) 81.
  
\bibitem{rebPLA87}
F.M.~Paiva, 
M.J.~Rebou\c{c}as 
and A.F.F.~Teixeira,
  ``Time travel in the homogeneous Som--Raychaudhuri universe,'' Phys. Lett. \textbf{A126} (1987) 168.

\bibitem{Leigh:2007wf} 
  R.G.~Leigh and A.C.~Petkou,
  ``Gravitational duality transformations on (A)dS$_4$,''
  JHEP {\bf 0711}, 079 (2007)
  [arXiv:0704.0531 [hep-th]].

\bibitem{Mansi:2008br} 
  D.S.~Mansi, A.C.~Petkou and G.~Tagliabue,
  ``Gravity in the $3+1$-split formalism I: holography as an initial value problem,''
  Class.\ Quant.\ Grav.\  {\bf 26} (2009) 045008
  [arXiv:0808.1212 [hep-th]].
  
\bibitem{Mansi:2008bs} 
  D.S.~Mansi, A.C.~Petkou and G.~Tagliabue,
  ``Gravity in the $3+1$-split formalism II: self-duality and the emergence of the gravitational Chern--Simons in the boundary,''
  Class.\ Quant.\ Grav.\  {\bf 26}, 045009 (2009)
  [arXiv:0808.1213 [hep-th]].

\bibitem{Caldarelli:1999xj}
  M.M.~Caldarelli, G.~Cognola and D.~Klemm,
  ``Thermodynamics of Kerr--Newman--AdS black holes and conformal field
  theories,''
  Class.\ Quant.\ Grav.\  {\bf 17} (2000) 399
  [arXiv:hep-th/9908022].
  
\bibitem{Gibbons:2004ai}
  G.W.~Gibbons, M.J.~Perry and C.N.~Pope,
  ``The first law of thermodynamics for Kerr--anti-de Sitter black holes,''
  Class.\ Quant.\ Grav.\  {\bf 22} (2005) 1503
  [arXiv:hep-th/0408217].
      
\bibitem{misner:1963}
C. Misner, ``The flatter regions of Newman, Unti and Tamburino's generalized Schwarzshild space,'' Jour. Math. Phys. \textbf{4} (1963) 924.
 \bibitem{bonnor:1969}
 W.B. Bonnor, ``A new interpretation of the NUT metric in general relativity,'' Proc. Camb. Phil. Soc. \textbf{66} (1975) 145.
 
 
\bibitem{dowker:1974}
J.S. Dowker, ``The NUT solution as a gravitational dyon,'' GRG \textbf{5} (1974) 603. 

\bibitem{Astefanesei:2004kn}
  D.~Astefanesei, R.B.~Mann and E.~Radu,
  ``Nut charged spacetimes and closed timelike curves on the boundary,''
  JHEP {\bf 0501} (2005) 049
  [arXiv:hep-th/0407110].

\bibitem{Chamblin:1998pz}
  A.~Chamblin, R.~Emparan, C.V.~Johnson and R.C.~Myers,
  ``Large $N$ phases, gravitational instantons and the nuts and bolts of AdS
 holography,''
  Phys.\ Rev.\ {\bf D59} (1999) 064010
  [arXiv:hep-th/9808177].

\bibitem{Caldarelli:2008pz}
  M.M.~Caldarelli, R.~Emparan and M.J.~Rodriguez,
  ``Black rings in (anti)-de Sitter space,''
  JHEP {\bf 0811} (2008) 011
  [arXiv:0806.1954 [hep-th]].

\bibitem{Mann:1996gj}
  R.B.~Mann,
  ``Pair production of topological anti-de Sitter black holes,''
  Class.\ Quant.\ Grav.\  {\bf 14} (1997) L109
  [gr-qc/9607071].

\bibitem{Vanzo:1997gw}
  L.~Vanzo,
  ``Black holes with unusual topology,''
  Phys.\ Rev.\ {\bf D56} (1997) 6475
  [gr-qc/9705004].
  
\bibitem{Brill:1997mf}
  D.R.~Brill, J.~Louko and P.~Peldan,
  ``Thermodynamics of $3+1$-dimensional black holes with toroidal or higher genus horizons,''
  Phys.\ Rev.\ {\bf D56} (1997) 3600
  [gr-qc/9705012].

\bibitem{Emp99}
R. Emparan, ``AdS/CFT duals of topological black holes and the entropy of zero energy states,'' JHEP \textbf{036} 1999 9906 [arXiv:hep-th/9906040].

\bibitem{Klemm:1997ea}
  D.~Klemm, V.~Moretti and L.~Vanzo,
  ``Rotating topological black holes,''
  Phys.\ Rev.\ {\bf D57} (1998) 6127
   [Erratum-ibid.\ {\bf D60} (1999) 109902]
  [arXiv:gr-qc/9710123].


\bibitem{demianski:1966}
M. Demia\'nski and E.T. Newman, ``A combined Kerr--NUT solution of the Einstein field equations,'' Bulletin de l'Acad\'emie Polonaise des Sciences, \textbf{XIV} (1966) 653.

\bibitem{Mazur:1986gb}
  P.O.~Mazur,
  ``Spinning cosmic strings and quantization of energy,''
  Phys.\ Rev.\ Lett.\  {\bf 57} (1986) 929.

\bibitem{Bhattacharyya:20082}
  S.~Bhattacharyya, R.~Loganayagam, I. Mandal, S.~Minwalla and A. Sharma,
  ``Conformal nonlinear fluid dynamics from gravity in arbitrary dimensions,''
  JHEP {\bf 0812} (2008) 116
  [arXiv:0809.4272 [hep-th]].

\bibitem{Plebanski:1976gy}
  J.F.~ Pleba\'nski and M.~Demia\'nski  ,
  ``Rotating, charged, and uniformly accelerating mass in general relativity,''
  Annals Phys.\  {\bf 98} (1976) 98.


\end{thebibliography}
\end{document}